\documentclass[]{spie}  %>>> use for US letter paper
\usepackage{amsmath,amsfonts,amssymb}
\usepackage{graphicx}
\usepackage[colorlinks=true, allcolors=blue]{hyperref}

\title{Characterization of a multi-etalon array for ultra-high resolution spectroscopy}

\author[a]{Surangkhana Rukdee}
\affil[a]{Center for Astrophysics ${\rm \mid}$ Harvard {\rm \&} Smithsonian 60 Garden Street Cambridge, MA, USA}

\author[a]{Sagi Ben-Ami}
\author[a]{Andrew Szentgyorgyi}
\author[a]{Mercedes López-Morales}
\author[a]{David Charbonneau}
\author[a]{Juliana García-Mejía}
\authorinfo{Further author information: (Send correspondence to S.R.)\\S.R.: E-mail: surangkhanar@gmail.com}

\begin{document} 
\maketitle

\begin{abstract}
% The upcoming Extremely Large Telescopes (ELTs) are expected to have enough collecting area to detect potential biosignature gases in the atmosphere of terrestrial planets around nearby stars. One of the most promising detection methods is through cross correlation of high-resolution transmission spectra. While current instruments typically achieve spectral resolution of 100,000, recent studies show that a spectral resolution of 300,000-400,000 is optimal to detect molecular oxygen (O2) in the atmosphere of an earth analog with the ELTs. Therefore, we implement an ultra-high spectral resolution booster to be coupled in front of a high resolution spectrograph to create a hyperfine chained spectral profile and fully sample the oxygen A-band (760 nm). We present results from our lab prototype, with a resolving power of 600,000. In addition, we discuss how dualons in our prototype reach a higher overall throughput than etalons.
The upcoming Extremely Large Telescopes (ELTs) are expected to have the collecting area required to detect potential biosignature gases in the atmosphere of rocky planets around nearby low-mass stars. Some efforts are currently focusing on searching for molecular oxygen (${\rm O_2}$), since ${\rm O_2}$ is a known biosignature on Earth. One of the most promising methods to search for ${\rm O_2}$ is transmission spectroscopy in which high-resolution spectroscopy is combined with cross-correlation techniques. In this method, high spectral resolution is required both to resolve the exoplanet's ${\rm O_2}$ lines and to separate them from foreground telluric absorption. While current astronomical spectrographs typically achieve a spectral resolution of 100,000, recent studies show that resolutions of 300,000 -- 400,000 are optimal to detect ${\rm O_2}$ in the atmosphere of earth analogs with the ELTs. Fabry Perot Interferometer (FPI) arrays have been proposed as a relatively low-cost way to reach these resolutions. In this paper, we present performance results for our 2-FPI array lab prototype, which reaches a resolving power of 600,000. We further discuss the use of multi-cavity etalons (dualons) to be resolution boosters for existing spectrographs.

\end{abstract}

% Include a list of keywords after the abstract 

\section{Introduction}
\label{sec:intro}  % \label{} allows reference to this section
%%%% THIS PART MAYBE TOO SIMPLE? (HIRES 101)%%%

{High resolution spectroscopy ($R\geqslant 100,000$) plays an essential role in astronomical observations, from detailed measurements of stellar chemical composition, which provides an exquisite probe of nucleosynthesis and stellar evolution \cite{2012A&A...541A.137M}, through studies of open cluster ages \cite{10.1093/mnras/stz3008}, to precise mass measurements of exoplanets \cite{1995Natur.378..355M}. In the last decade or so, the use of high resolution spectroscopy has also extended to observations of gas giant exoplanet atmospheres {\cite{2002PASP..114..826B} \cite{Redfield_2008}}, with several detections of carbon oxide and water in transiting and non-transiting planets {\cite{2010Natur.465.1049S} \cite{2012ApJ...753L..25R} \cite{2018arXiv180604617B}}. Snellen et al.\cite{2010Natur.465.1049S} \cite{2013ApJ...764..182S} suggested that high resolution spectroscopy could also be used to search for potential biomarkers in Earth analogs transiting around nearby stars. In particular they showed that molecular oxygen (${\rm O_2}$) could be detected using upcoming $R = 100,000$ spectrographs on Extremely Large Telescopes (ELTs), although even with the large collecting areas of ELTs it will still be necessary to observe dozens of transits to detect a signal \cite{Rodler_2014}. Although at $R = 100,000$ ${\rm O_2}$ lines from the atmosphere of an Earth analog are not fully resolved\cite{Ben_Ami_2018}, a recent study \cite{2019BAAS...51c.162L} argues that $R = 300,000-400,000$ will enable us to fully resolve ${\rm O_2}$, thereby reducing the number of required transits by about 34\%.}

It is difficult to reach $R > 100,000$ with traditional seeing-limited high-resolution echelle spectrographs designed for 6.5-10 m class telescopes: the resulting beam size would require that a very large and prohibitively expensive echelle grating be employed. Several techniques have been suggested to bypass this problem. One such technique involves the use of an externally dispersed spectrograph to achieve very high resolution, while a smaller standard echelle acts as an order separator. The Dispersed Fixed Delay Interferometer (DFDI)\cite{Ge_2006},\cite{2002PASP..114.1016G} and the TripleSpec Exoplanet Discovery Instrument (TEDI) \cite{10.1117/12.735474} employ such a technique to carry out Doppler exoplanet surveys. DFDI optimizes the optical delay in the interferometer and reduces photon noise by measuring multiple fringes over a broad band. The interferometer is then coupled with a low- to medium-resolution postdisperser \cite{2002ApJ...571L.165G}. TEDI has an Externally Dispersed Interferometer (EDI) coupled with a conventional high resolution R=20,000 spectrograph. The prototype demonstrated a factor of six increase in resolving power\cite{10.1117/12.735474}, but at the cost of increased exposure time since the EDI make separate measurements at numerous values of the fixed delay. 

Ben-Ami et al. (2018) (hereafter B18) suggested a Fabry Perot Interferometer (FPI) array design that, combined with a standard high resolution spectrograph such as the GMT-Consortium Large Earth Finder (G-CLEF) \cite{2016SPIE.9908E..22S}, could reach the high resolutions and signal levels needed for the ${\rm O_2}$ science case described above. B18 envision an array of eight etalons that can fully sample the spectral profile of the ${\rm O_2}$ A-band. In addition to increasing the resolution using that FPI array, B18 also highlight how {\it dualons} (multi-mirror etalons) might be able to further increase the instrument's throughput and reduce the amount of telescope time necessary to achieve a detection.

In this paper we present the results of tests for a 2-FPI array prototype built in the lab to test the instrument concept proposed by Ben-Ami et al. (2018). In Section 2 we describe our prototype. In Section 3 we describe the properties of etalons and dualons and detail the model predictions for the specific parameters of our setup. Section 4 describes the tests performed with the prototype. We summarize and discuss our test results in Section 5.

\section{Properties of Fabry Perot Interferometers} \label{sec:properties}
\subsection{Etalon}

\begin{figure}[h!]
\center\includegraphics[width=.3\textwidth]{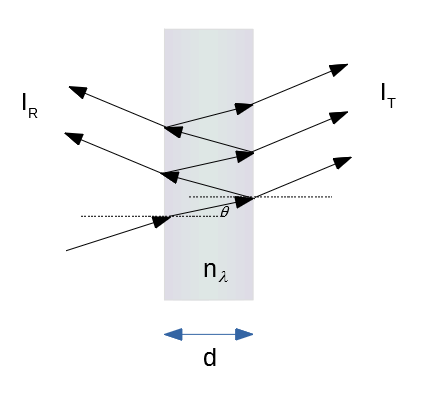}
\caption{The light path inside etalon. $I_R$ is the intensity of reflection. $I_T$ is the intensity of transmission. $n_{\lambda}$ is the refractive index. $d$ is the separation between the two surfaces.  \label{fig:etalon_explanation}}
\end{figure}

A plano-plano Fabry Perot etalon consists of two parallel flat surfaces separated by a small gap. The light path inside the etalon (Figure \ref{fig:etalon_explanation}) yields transmission and reflection intensities given by\cite{vaughan}:

\begin{equation}
I_T = \frac{T^2}{(1 - R)^2(1 + F \sin^2 (\frac{\phi}{2}))}
\label{eq:I_T}
\end{equation}

\begin{equation}
I_R = \frac{F \sin^2 (\frac{\phi}{2})}{1 + F \sin^2 (\frac{\phi}{2})}
\label{eq:I_R}
\end{equation}

\noindent
where $T$ and $R$ are the surface intensity reflection and transmission coefficients, respectively. The finesse $F$ is defined by $F = 4R/(1 - R)^2$. It describes the resolution of the instrument. The separation between profile peaks of each FPI would then be the free spectral range (FSR). The phase lag $(\phi = \frac{2\pi}{\lambda} 2dn_\lambda\cos\theta)$ is the phase difference caused by the optical delay for successive reflections. $\phi$ depends on the separation of the reflective surfaces d, the angle of incidence $\theta$, and the wavelength dependent refractive index $n_{\lambda}$. The latter is set by the dispersion formula of fused silica \cite{Malitson:65}.

%\begin{equation}
%n_{\lambda}^2-1=\frac{0.6961663\lambda^2}{\lambda^2-0.0684043^2}+\frac{0.4079426\lambda^2}{\lambda^2-0.1162414^2}+\frac{0.8974794\lambda^2}{\lambda^2-9.896161^2}
%\label{eq:dispersion formula}
%\end{equation}

\begin{table}[h!]
\renewcommand{\thetable}{\arabic{table}}
\centering
\caption{Etalon Specification} \label{tab:etalon spec}
\begin{tabular}{ll}
\hline
\hline
%\decimals
Manufacturer &  Light Machinery  \\
Product Number &  OP-9689  \\
Clear Aperture &  13 nm  \\
Thinkness uniformity &  1.2 nm  RMS   \\
Absolute thickness &  26.308 $\pm$ 0.05 mm    \\
Dimension &   25.36  mm   \\
Reflectivity &   72.5\%  \\
%Calculated finesse &   9  \\
Coating &   750-780 nm  \\
Reference edge square to mirrors &   2.9'  \\
\hline
\end{tabular}
\end{table}

The specifications of one etalon used in this work are summarized in Table. \ref{tab:etalon spec}. The absolute thickness between the two surfaces is 26.308 mm, while in the second etalon it is 130 nm further separated. We chain two etalons given both the transmission and reflection properties of the etalon to create an ultra high resolution spectral comb-like profile \cite{Ben_Ami_2018}.
\subsection{Dualon (Multi-Mirrors)} \label{sec:multi-mirrors}

\begin{figure}[h]
\begin{center}
\includegraphics[width=0.5\textwidth]{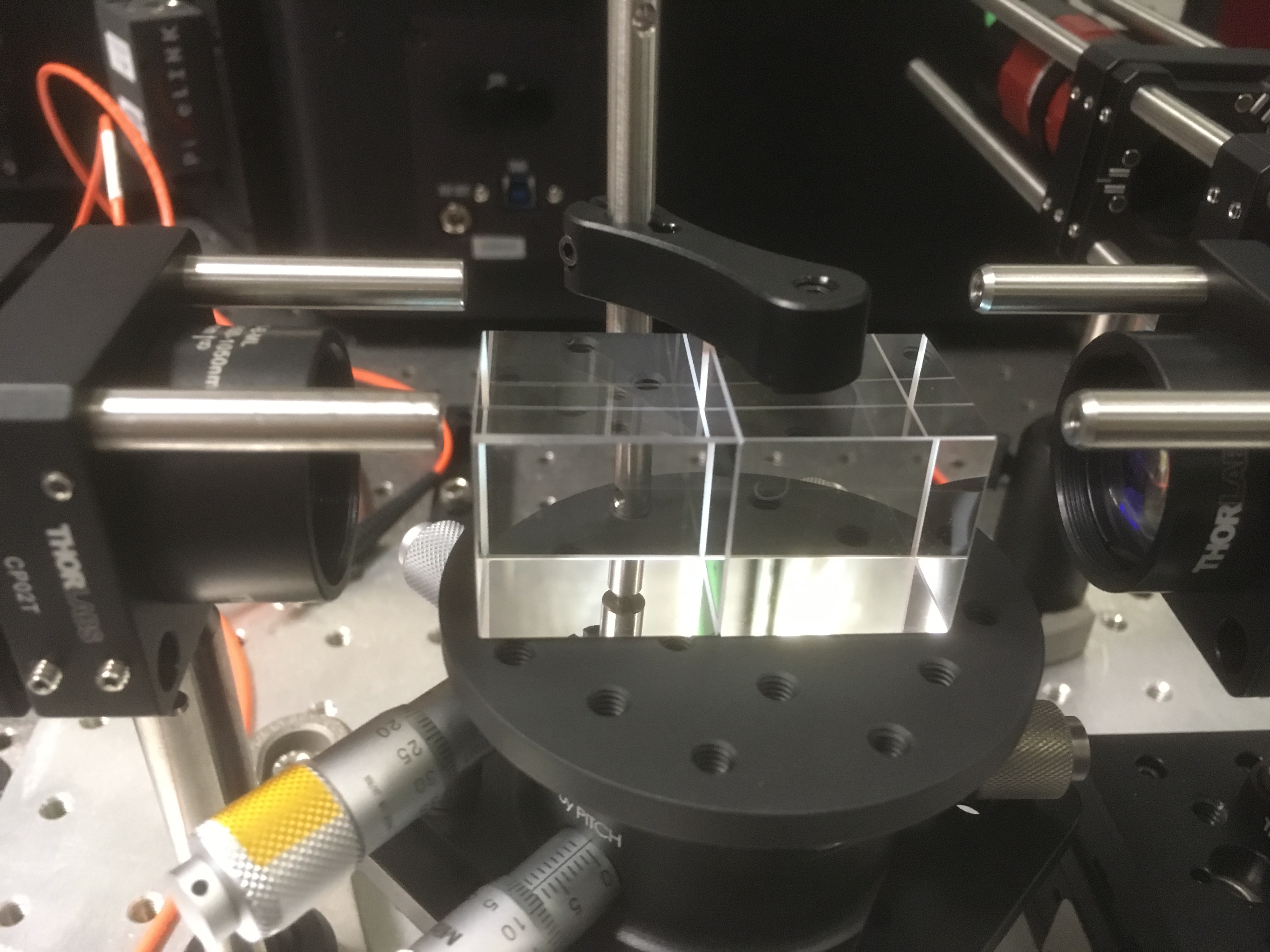}
\caption{One of the two dualons in the prototype. It consists of two etalon attached to each other. In this case the light will come from the right hand side of the image. Part of the light transmits ($I_T$) to the focusing lens to the left hand side and part of the light reflects back ($I_R$) to the right hand side.
\label{fig:dualon}}
\end{center}
\end{figure}

A dualon is a double-etalon or multi-surface etalon (Figure. \ref{fig:dualon}). It consists of three reflective surfaces: the two outer ones, and an inner one. This results in a much higher contrast than a standard etalon for a given finesse. For $N$ reflective surfaces, the transmitted intensity through a dualon is expressed as \cite{vandeStadt:85}:

\begin{equation}
I_T = \frac{T^{(2N-2)}}{(1 - R)^{(2N-2)}(1 + F \sin^{(2N-2)} (\frac{\phi}{2}))}
\label{eq:I_T_multimirror}
\end{equation}

\noindent
where $T$ and $R$ are the surface intensity reflection and transmission coefficients, respectively. N is the number of mirrors. The finesse F is given by: 

\begin{equation}
F = \frac{4R}{(1 - R)^{(2N-2)}}
\label{eq:finesse_multimirror}
\end{equation}

\begin{figure}[h]
\begin{center}
\includegraphics[width=0.6\textwidth]{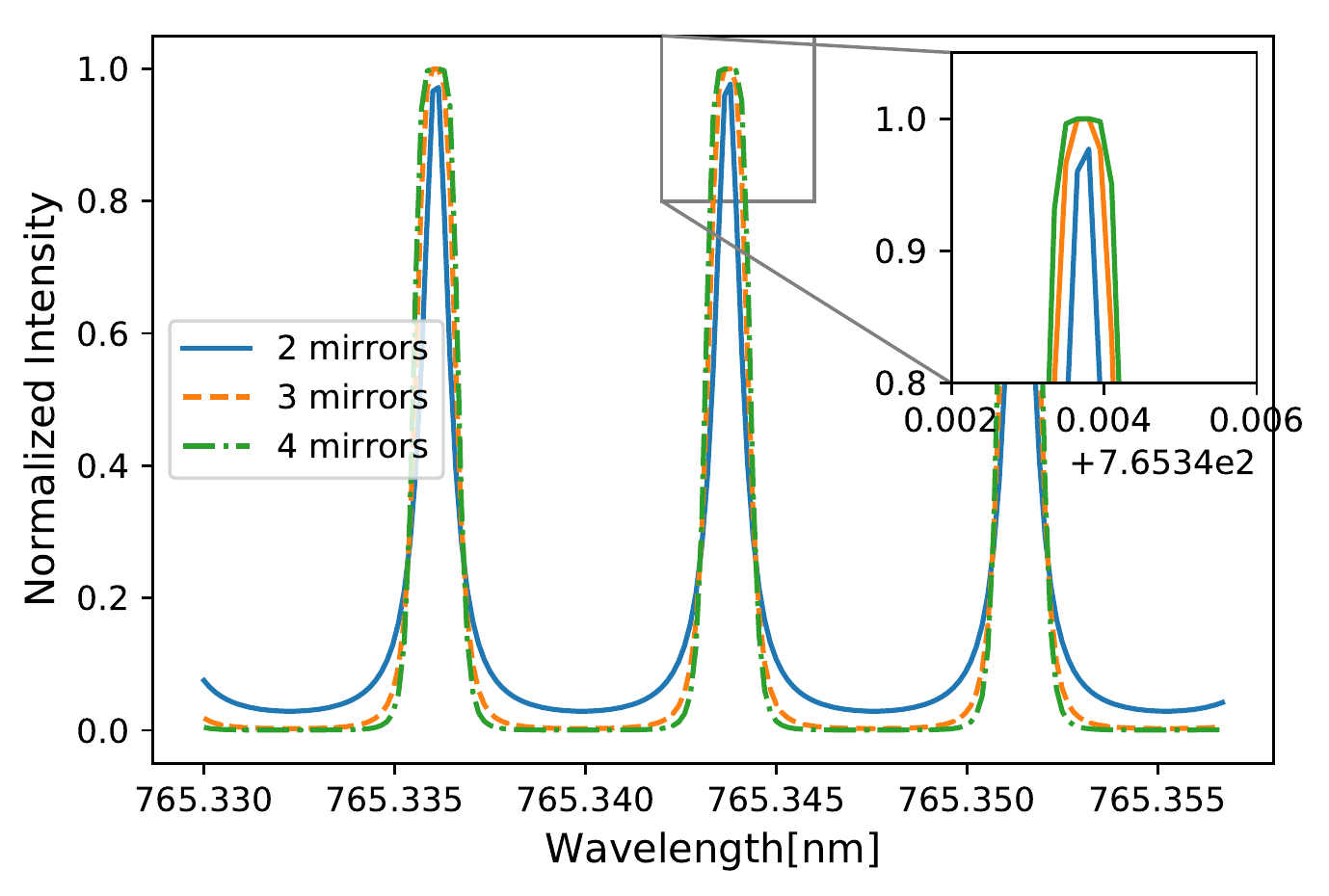}
\caption{Theoretical model of a transmission profile from multi surface etalon. 
\label{fig:transmission_multi_surface}}
\end{center}
\end{figure}

 We employ Eq. \ref{eq:I_T_multimirror} to generate the transmission profile of a FPI with different number of mirrors. The expected profiles for FPIs with 2, 3, and 4 mirrors are shown in blue, orange and green in Figure \ref{fig:transmission_multi_surface}. Increasing the number of reflection surfaces of an FPI improves the contrast of the transmission profile.

\begin{figure}[h]
\begin{center}
\includegraphics[width=0.6\textwidth]{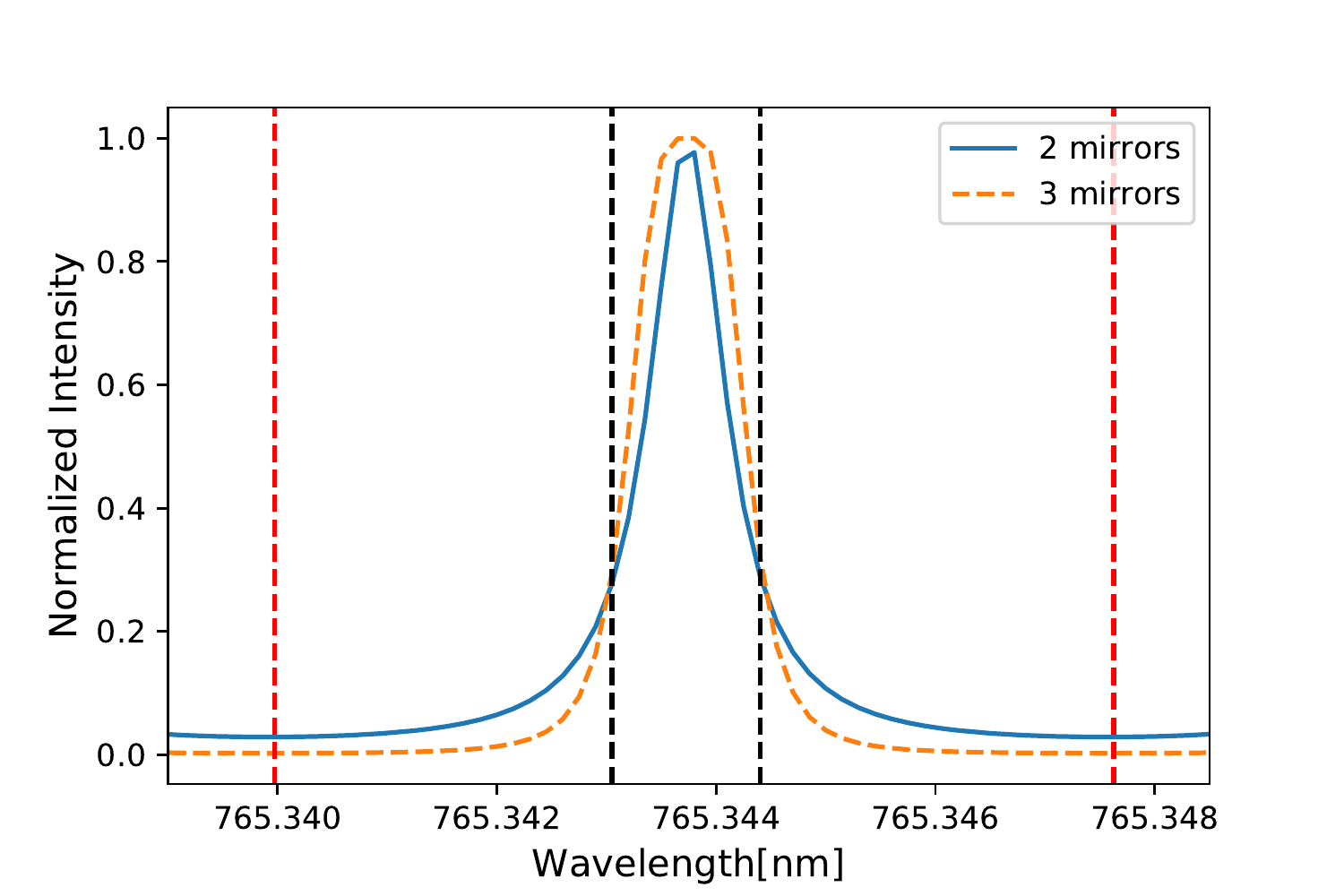}
\caption{Transmission profile of a multi surface FPI with 2 mirrors in solid blue line and 3 mirrors in dashed orange line. The vertical dashed lines show the total integration interval (red) and the interval at 30\% intensity (black).
\label{fig:Purity}}
\end{center}
\end{figure}

We compare the difference of the light loss between an etalon and a dualon by means of the purity ratio (Figure \ref{fig:Purity}). The purity ratio is the ratio between the integrated area between the two valleys (red cut) and the integrated area of the peak of each profile (black cut) where the cut is at the normalised intensity level of 0.3. The model comparison is considering the ideal case without beam degradation. We obtain a purity of 61.80\% from the etalon (2 mirrors) and 88.95\% from the dualon (3 mirrors).

% Juliana rewrite suggestion: Referring to Figure 4, we define the purity ratio as the ratio of the integrated area under a transmission peak to the area under the same peak \textit{and} its valleys. The integrated area under the transmission peak is bounded by the vertical lines intersecting the transmission peak at a normalized intensity of 0.3 (dashed black lines in Fig. 4). 

% also I would add a sentence explaining how you choose the place of the red vertical lines. 

\subsection{Model Prediction} \label{sec:prediction}
\begin{figure}[h!]
\centerline{\includegraphics[width=.5\textwidth]{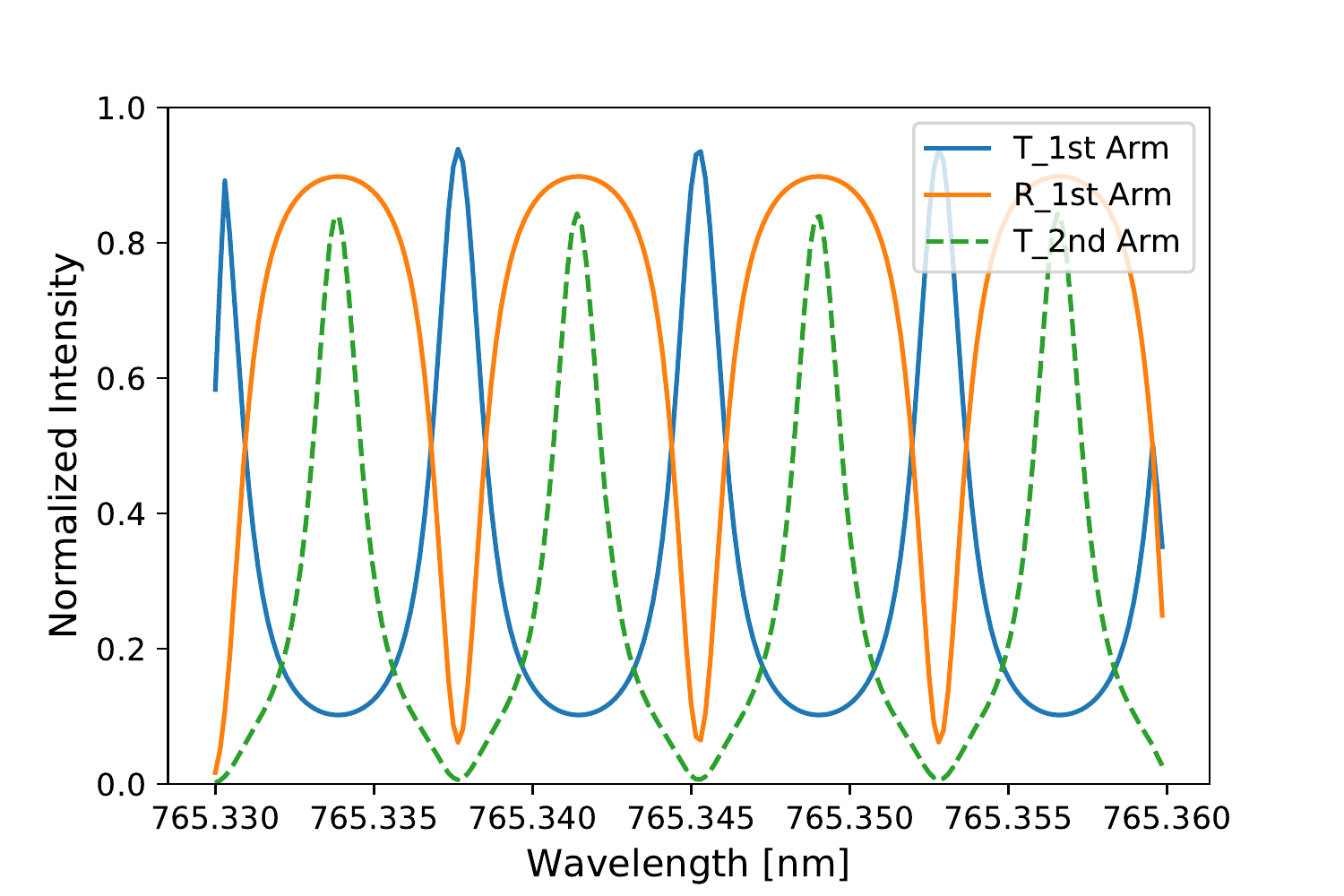}}
\caption{The transmission (T) profiles of the first arm (blue solid line) and the second arm (green dashed line) and the reflection (R) profile (orange solid line) for a single mode fiber with incident angle, R=0.71, and finess = 8.8. \label{fig:TRsmf}}
\end{figure}

An idealized FPI can achieve a very high resolution through its comb-like throughput. However, there are many parameters that could affect the performance in the system. The spatial and temporal intensity modulation across the input light source and non-uniform illumination of the interferometer can degrade its performance. In terms of alignment, poor focusing of the collimating and focusing lenses, and misalignment alter the performance of the system. Thus, we implement a numerical simulation of the transmission model (Eq \ref{eq:I_T}) to predict the output signal from each arm from the setup in Figure \ref{fig:prototype}. The reflected wavelengths from the first arm can be explained in the Eq \ref{eq:I_R}. The transmission and reflection profiles of two etalon arrays (arms) are shown in Figure \ref{fig:TRsmf}.  The reflected light from the first etalon, set to a position with a small angle to hit the tip of the prism and goes into the second arm. For this chained FPI, the reflected signal from the first arm become the input signal for the second arm. Thus, the transmitted signal from the second arm can be explained as: 

%% JGM Comments about paragraph above: what do you mean by very sensitive measurement in the first sentence? do you mean high throughput at quasi-discrete wavelengths?

\begin{equation}
I_{T_2} = I_{R_1} \frac {T_1^2}{(1-R_1)^2(1 + F \sin^2 (\frac{\phi}{2}))} 
\label{eq:I_T2}
\end{equation}

We consider the degradation due to an imperfect collimation and take into account a finite divergence in the incident angle $\theta$ of the collimated beam as $\delta\theta_{max} = \pm\Phi_{fiber}/2f_{col}$ where $\Phi_{fiber}$ is the fiber diameter and $f_{col}$ is the focal length of the collimator. This contributes the phase lag term \cite{Ben_Ami_2018} of:

\begin{equation}
\delta\phi_{max} = \frac{4\pi}{\lambda}dn_\lambda\ (\sin(\theta\pm\frac{\Phi_{fiber}}{2f_{col}})-\sin\theta)
\label{eq:beam deviation}
\end{equation}

\begin{table}[h!]
\renewcommand{\thetable}{\arabic{table}}
\centering
\caption{Finesse budget}  \label{tab:finesse budget}
\begin{tabular}{cccc}
Finesse & Budget & Requirements \\
\hline
\hline
%\decimals
Reflectivity & 9.13  &  R = 71\% \\
Divergence &  0.02 &  $\Phi$ = 1.4 mrad\\
Parallelism &  38.27 &  $\Delta$ = 10 nm\\
Defect &  180.40 &  $\delta$ = 3 nm\\
\hline
Effective &  8.8 &  &\\
\hline
\end{tabular}
\end{table}

The finesse of a real system is a combination of several parts. This includes reflectivity, divergence, parallelism and defect. We obtain these values from the manufacturer. The individual contributions are shown in Table.2. These are combined to an effective finesse $F_e$ in Equation \ref{eq:Fe} described in the previous study \cite{2017Cersullo} about this effect. This includes reflectivity finesse $F_R = \pi \sqrt{R} / (1-R)$, divergence finesse  $F_\Phi = 4\lambda/ (\Phi^2d)$, parallelism finesse $F_P = \lambda/ 2\Delta$ and defect finesse $F_D = \lambda/ \sqrt2\delta$. Our budget is shown in Table. \ref{tab:finesse budget}. 

\begin{equation}
F_e = \sqrt{{\frac{1}{F_R^2}}+{\frac{1}{F_\Phi^2}}+{\frac{1}{F_P^2}}+{\frac{1}{F_D^2}}}
\label{eq:Fe}
\end{equation}

\section{Prototype} \label{sec:prototype}

In this study we set out to test a 2-FPI prototype as a simplified version of the 8-FPI array proposed by Ref.  \citenum{Ben_Ami_2018}. Our prototype is optimized for the {\rm O$_2$} A-band centered around 765 nm. We aim to demonstrate that the expected resolution and the instrument efficiency described in Section \ref{sec:prediction} can be achieved. Figures 6 and 7 show the schematic layout and laboratory setup of the prototype. The setup consists of a collimating lens, a FPI and a focusing lens, which we describe in detail below.

%\subsection{Optical Design}
\begin{figure}[h]
\centerline{\includegraphics[width=0.7\columnwidth]{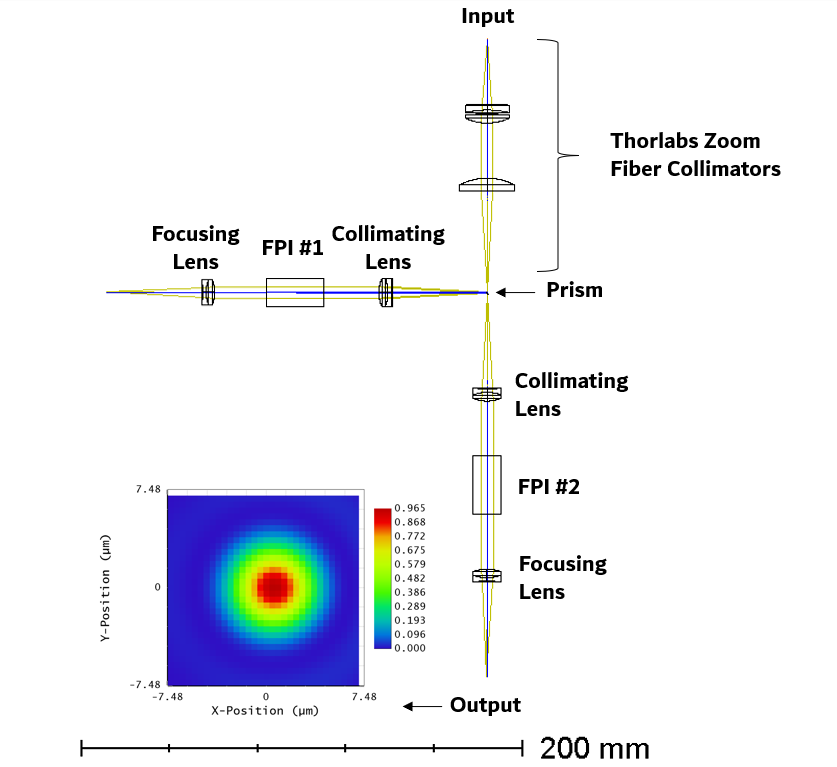}}
\caption{The optical design of our system of two FPIs where the light from the first FPI reflects into the second FPI. The output image from the focusing lens is displayed next to the optical layout. The box size represents a pixel size on the detector. The colorbar shows the Strehl Ratio.}
\label{fig:opitcal_design}
\end{figure}

The optical system design is illustrated in Figure \ref{fig:opitcal_design}: light enters the protoype through a  fiber collimator at the input fiber injection before hitting the prism very close to its vertex. The beam then proceeds towards  the first collimating lens and FPI\#1. Part of the beam is  transmitted through the FPI, and part of the light reflects back to the prism before going into the second collimating lens and FPI\#2. The transmitted, focused light is shown as output (Huygens Point Spread Function) within a pixel on the detector to the left of the optical design. The colorbar represents the image quality in terms of Strehl Ratio.

\begin{figure}[h!]
\centerline{\includegraphics[width=0.7\columnwidth]{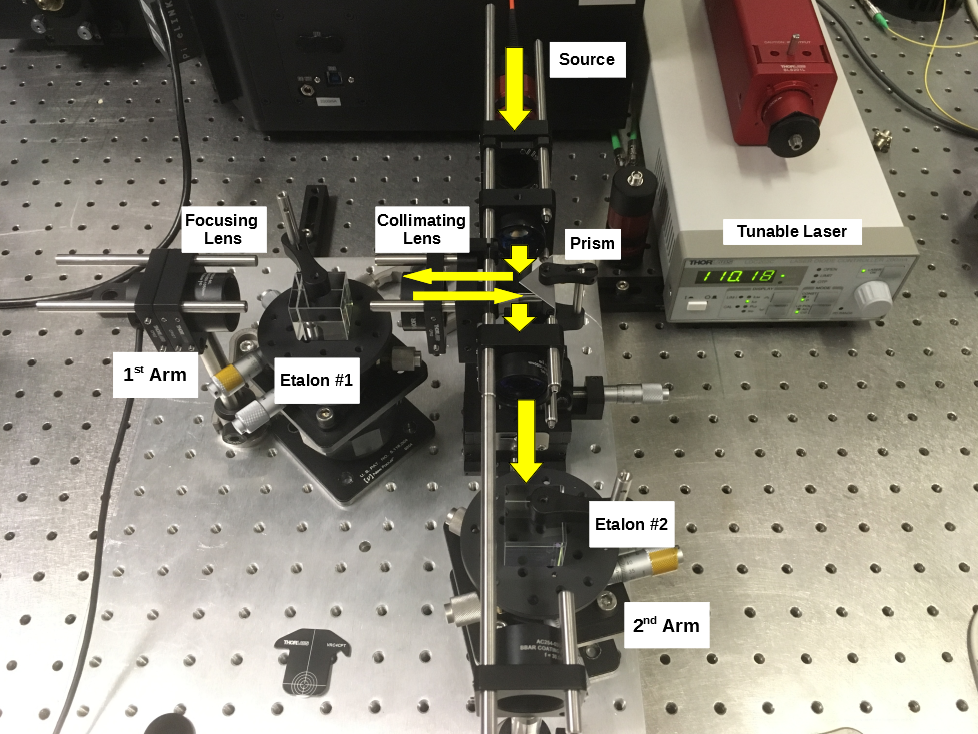}}
\caption{The Fabry Perot based Instrument for Oxygen Searches (FIOS) prototype setup consisting of two arms containing an etalon in each.}
\label{fig:prototype}
\end{figure}

We built the two-etalon prototype of the Fabry Perot Interferometer array  using commercially available components to test the instrument concept described in Ref. \citenum{Ben_Ami_2018}. Our etalon specifications is shown in Table \ref{tab:etalon spec}. We use two etalon cubes, which are identical in size but have slightly different separation (d). Figure \ref{fig:prototype} shows our implementation of the two etalon array. The light from a tunable laser is collimated and focused onto a knife-edge prism. The beam is reflected into the first arm (to the left), which consists of a collimating lens, an etalon and a focusing lens. Wavelengths that satisfy the etalon phase condition are transmitted. Other wavelengths are reflected towards the prism second reflective facet and go into a second arm with an identical setup. We place a photodiode sensor, connected to a power-meter, at the image plane of each arm and obtain the data from both arms simultaneously. We measure the output signal in narrow wavelength ranges by stepping through wavelength space in 1.5 pm intervals. 

\newpage
\section{Results} \label{sec:results}
\subsection{Beam Deviation} \label{sec:deviation}

Using different sizes of the fiber input alters the profile shape of the spectral feature. We investigate how the choice of fiber, namely a single mode or a multi-mode 50$\mu$m fiber, affects the spectral resolution of the multi-arm etalon. Feeding the system with different fiber sizes can introduce different deviations to the system since we keep the same focal length of the collimated beam using the same lens. This deviation is taken into account for the output signal model according to Equation (\ref{eq:beam deviation}).

\begin{figure}[h!]
\center
\includegraphics[width=.5\textwidth]{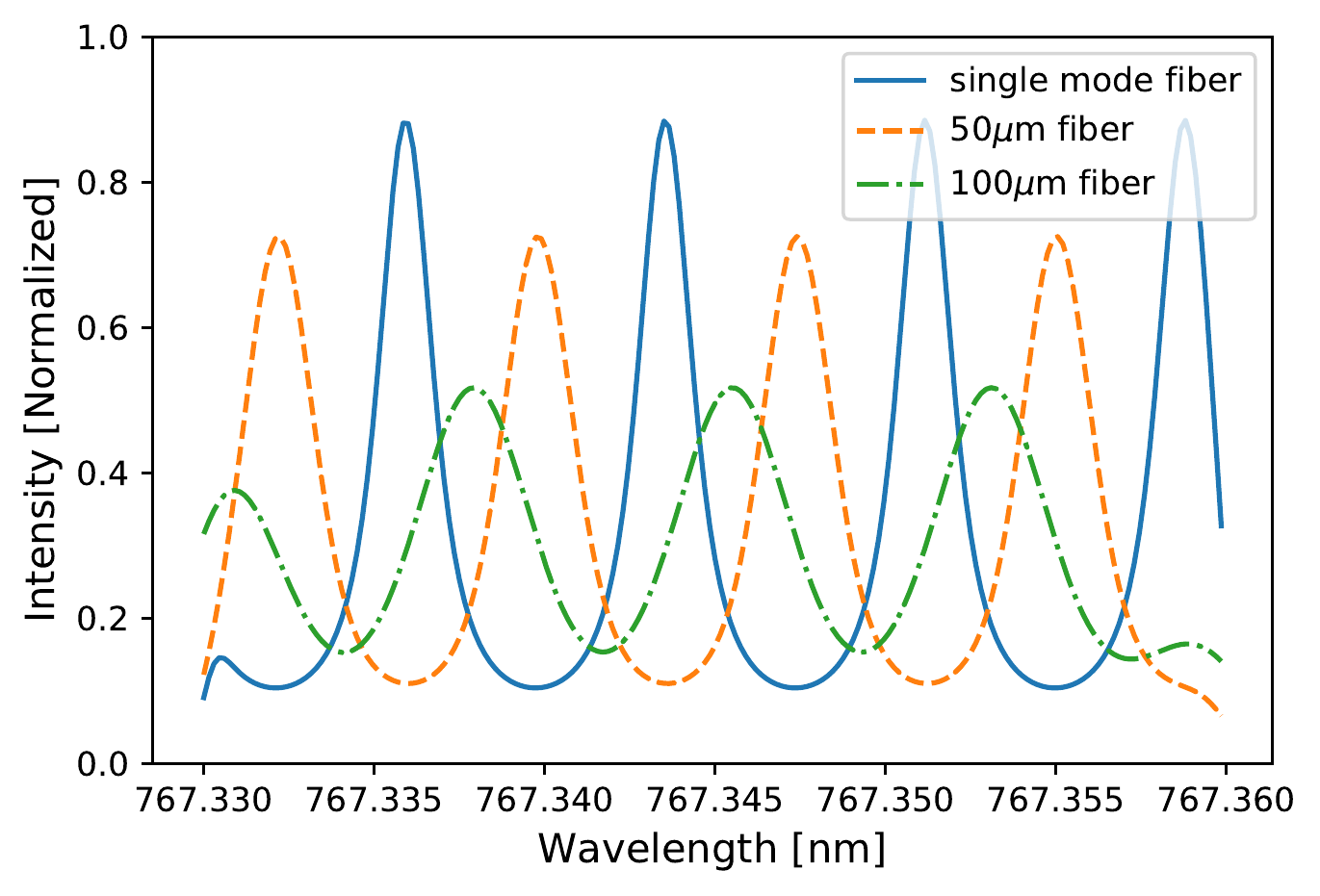}\hfill
\caption{The prediction of the deviation from using different fiber sizes and imperfectly collimated beam. The blue solid line shows the profile and intensity from the single mode fiber yielding the spectral resolution of R = 431,850 and throughput of 80\%. The orange dash line shows the case of a 50 $\mu$m fiber having R = 272,747 and a throughput of 50 \%. Green dash line is the 100 $\mu$m fiber of R = 178,696 and a throughput of 30\%.}
\label{fig:predict}
\end{figure}

Figure \ref{fig:predict} employs Equations \ref{eq:I_T} and \ref{eq:beam deviation} to demonstrate the impact of fiber sizes on spectral resolution. The spectral resolution $R$ is measured based on the full width half maximum (FWHM).
We find that using a single-mode (represented as 10 $\mu$m) fiber, a 50 $\mu$m fiber, and a 100 $\mu$m fiber yield spectral resolutions of $R = 431850$, $272747$, and $178696$,  respectively. In terms of efficiency, each increasing fiber size decreases the efficiency by $20\%$. In general the FPI achieves maximum efficiency when the light beam hits the FPI surface perpendicularly. Ideally the angle $\theta$ should be zero. However, our application employs both transmission and reflection properties. We need to direct the reflected light from the first arm into the second arm. Thus, a small angle is needed to enable that function although it decreases the total efficiency of the FPI. The location of each peak strongly depends on the initial wavelength and the initial angle of incidence $\theta$. In Figure \ref{fig:predict} we set the initial wavelength at 777.33 nm. We assume the initial angle according to the observed deviated beam size. However, we have the same system setup where we use the same focal length of the collimator ($f_{col}$ in Eq \ref{eq:beam deviation}) for all fiber sizes. We can see that the peak from each fiber size is not falling in the same place despite the same initial wavelength. The shift of these peaks is also observed in a previous study of the etalon's profile \cite{2017Cersullo}. Thus, we conclude that increasing fiber diameter or finite divergence introduces a loss of finesse and a shift in wavelength of the transmission peaks. This is due to the increasing number of rays that have the incidence angle $\theta > 0$.

\subsection{Etalons}

\begin{figure}[h!]
\center
\includegraphics[width=.45\textwidth]
{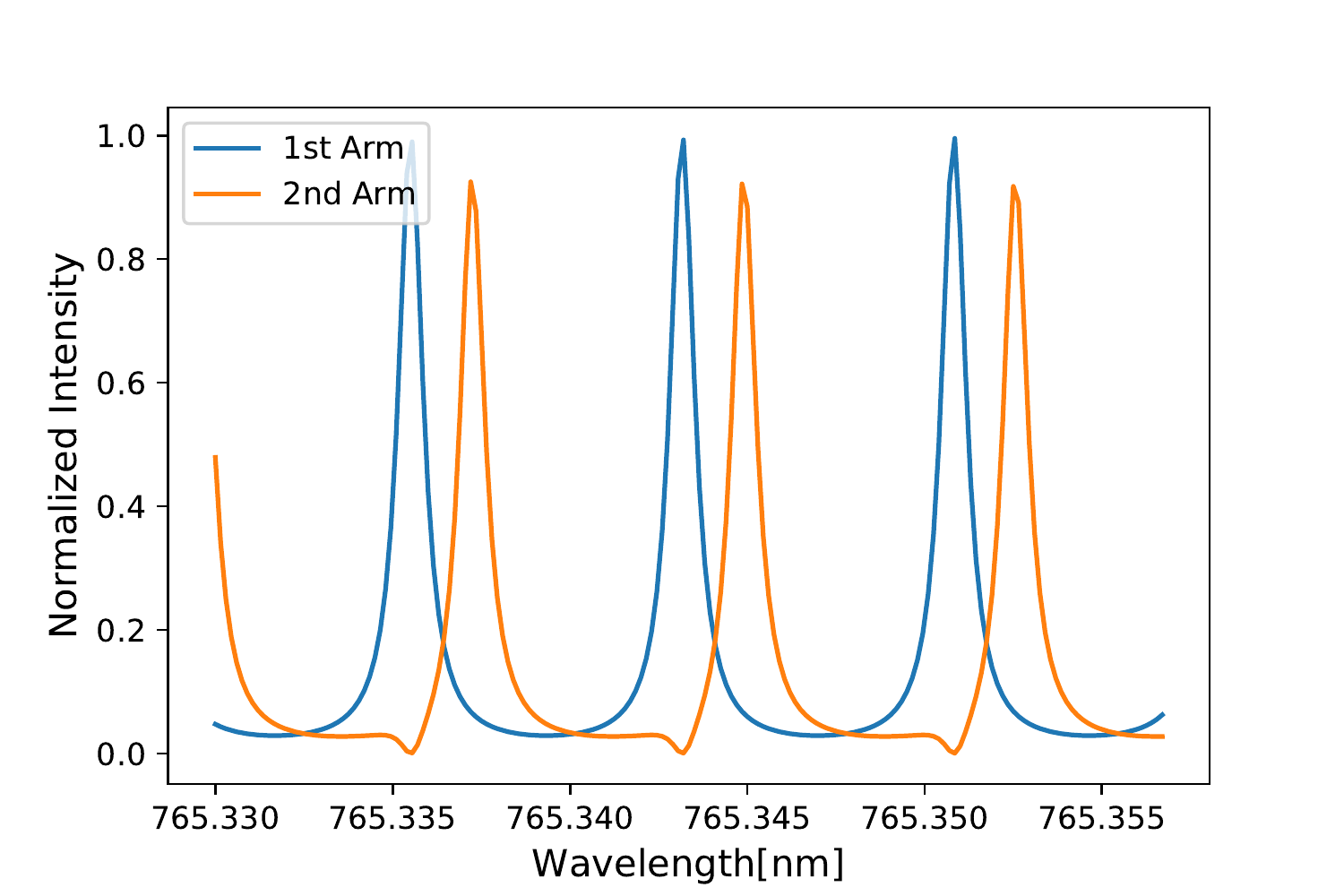}
\includegraphics[width=.45\textwidth]
{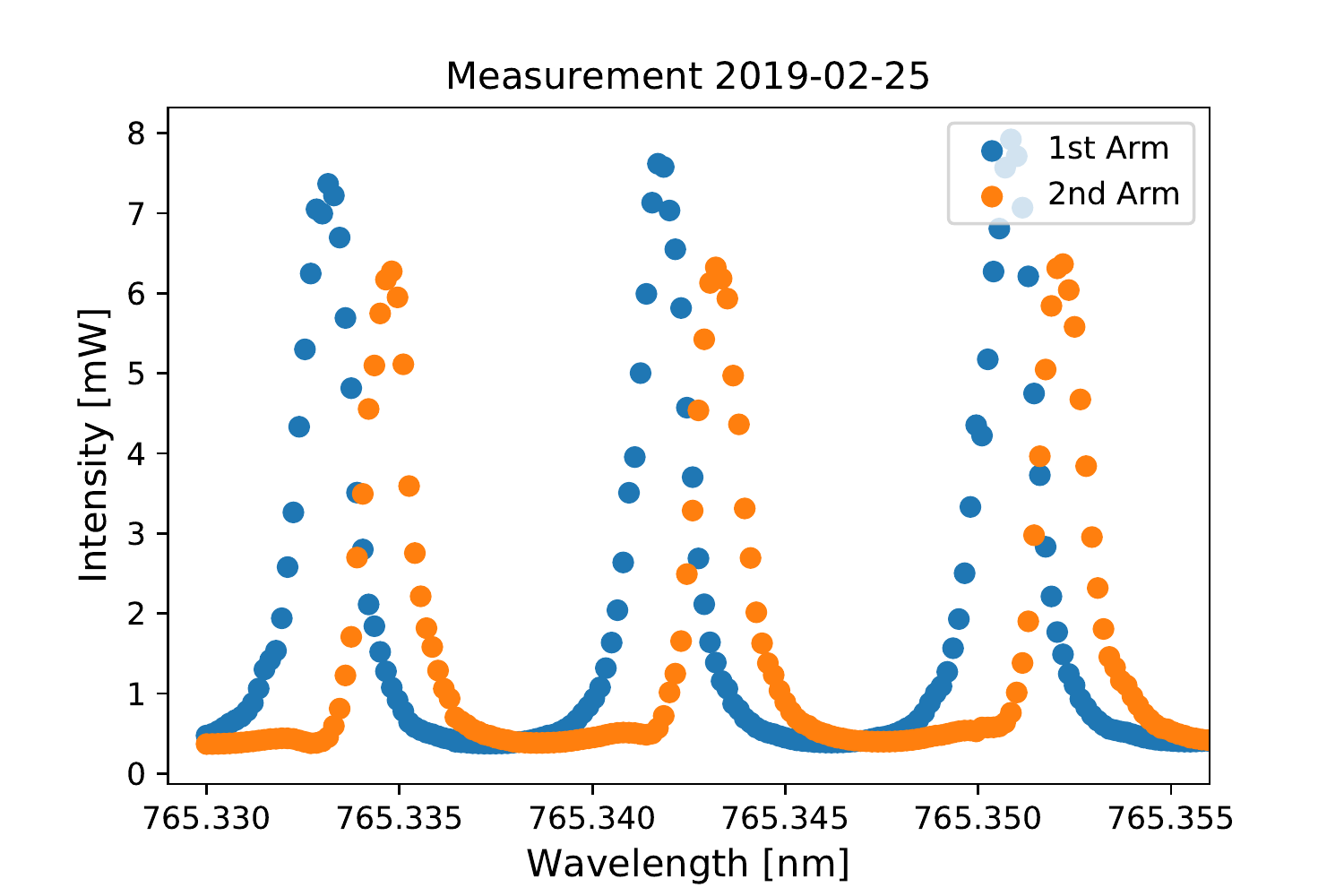}
\caption{The left panel shows the prediction of an ideal output beam from different arms in the case of single mode fiber, the reflectivity (R) = 0.71 and the finesse (F)=33.77. The right panel is the observation of the one resolution element setup of a single mode fiber with two etalons having a separation distance of 130 nm difference with the specifications in Table \ref{tab:etalon spec}.}
\label{fig:ideal1res}
\end{figure}

Figure \ref{fig:ideal1res} shows the ideal case of the setup with a high effective finesse for each etalon. 
%The system is fed by a single-mode fiber with a small deviation. 
The system is tuned so that the peaks in the second arm are shifted by one FWHM (R = 500,000 at 765 nm) compared to the first arm. This is possible because the etalons differ in thickness.
The achieved profile observed on the right side is similar to the simulation on the left side of Figure \ref{fig:ideal1res} due to adjustment of the tip-tilt stage, and small deviation from the single mode fiber.

%Our etalons differ in thickness and cause the distance between the transmission peaks to differ by R = 500,000 at 765 nm. 
%In the case that we keep the incident angles identical, the chained FPI creates continuous spectral features that allow us to sample the full ${\rm O_2}$ A band (760-780 nm). 
%Figure \ref{fig:ideal1res} shows the ideal case of this setup with a high effective finesse for each etalon. 

\begin{figure}[h!]
\center
\includegraphics[width=.45\textwidth]{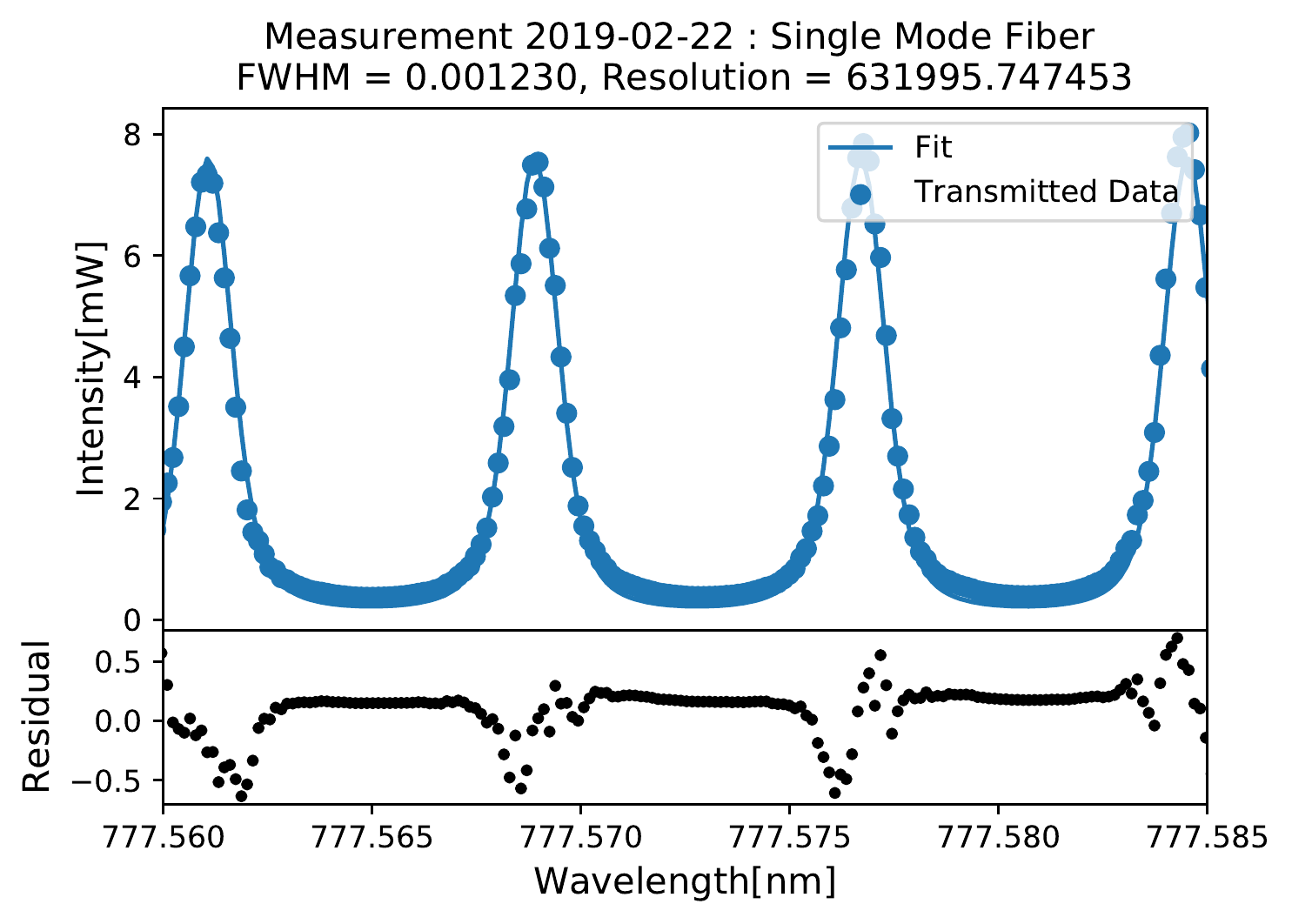}
\includegraphics[width=.45\textwidth]{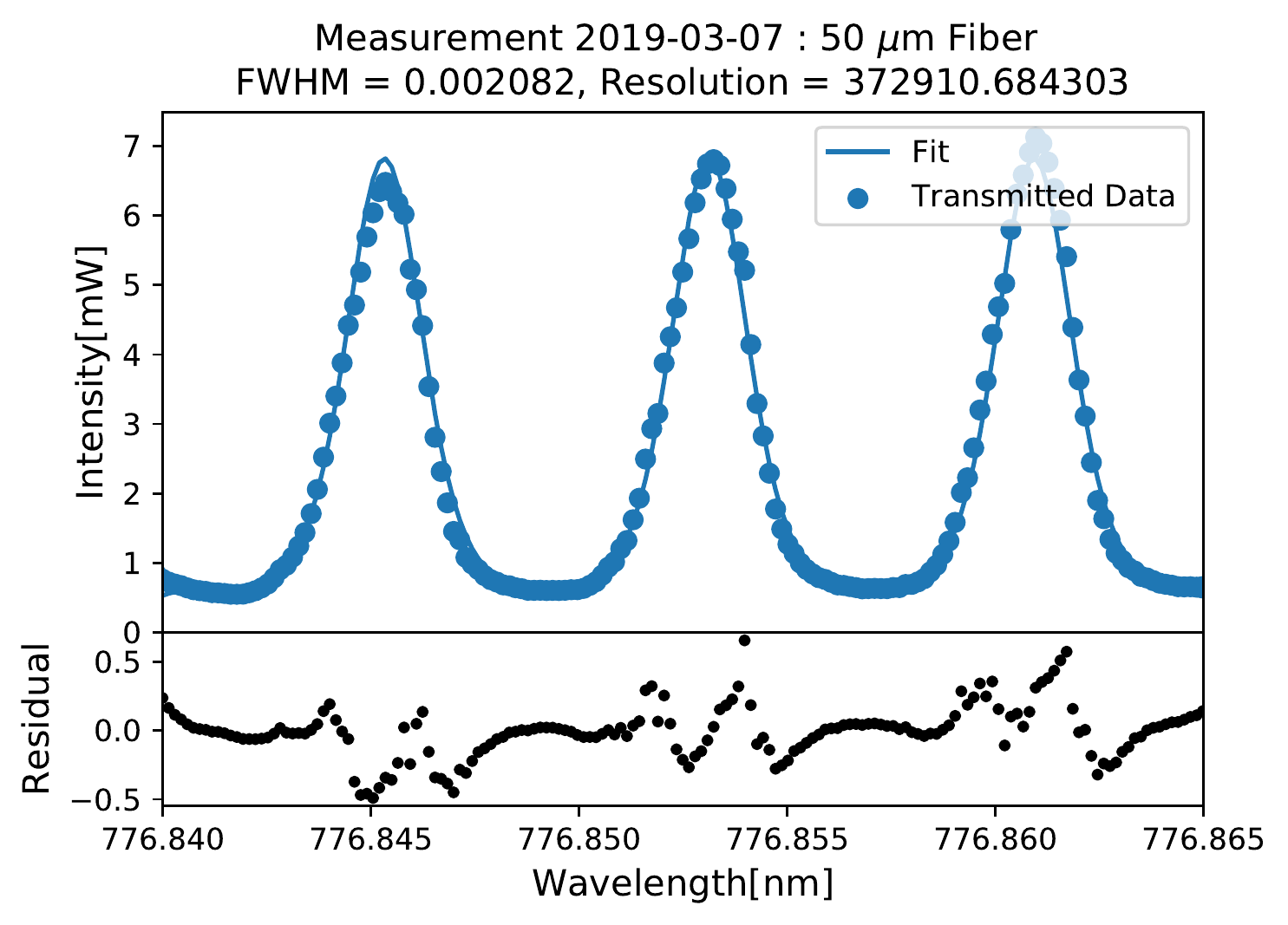}
\caption{Transmitted signal from the first arm fitted with Gaussian convolution. The single mode fiber (left) shows the feature of spectral resolution of R=631,995 while the 50 $\mu$m fiber (right) yields R=372,910.}
\label{fig:observed_onearm}
\end{figure}

We observed the deviation and the profile imperfection from the increasing fiber size as shown in Figure \ref{fig:observed_onearm}, agreeing with the prediction in Section \ref{sec:prediction}. In this figure, we show the true signal level observed from the prototype. We can see that the fiber size deviation also affects the throughput of the signal. We apply a $\chi^2$ fit to the model in Eq \ref{eq:I_T} for the parameters $\theta$, R, d, and $\lambda$. We account for various effects such as plate imperfections and the finite dispersion of the input beam by convolving the transmission signal with a Gaussian window, $w(n)=e^{-\frac{1}{2}(\frac{n}{\sigma})^2}$. 

\begin{figure}[h!]
\includegraphics[width=.47\textwidth]{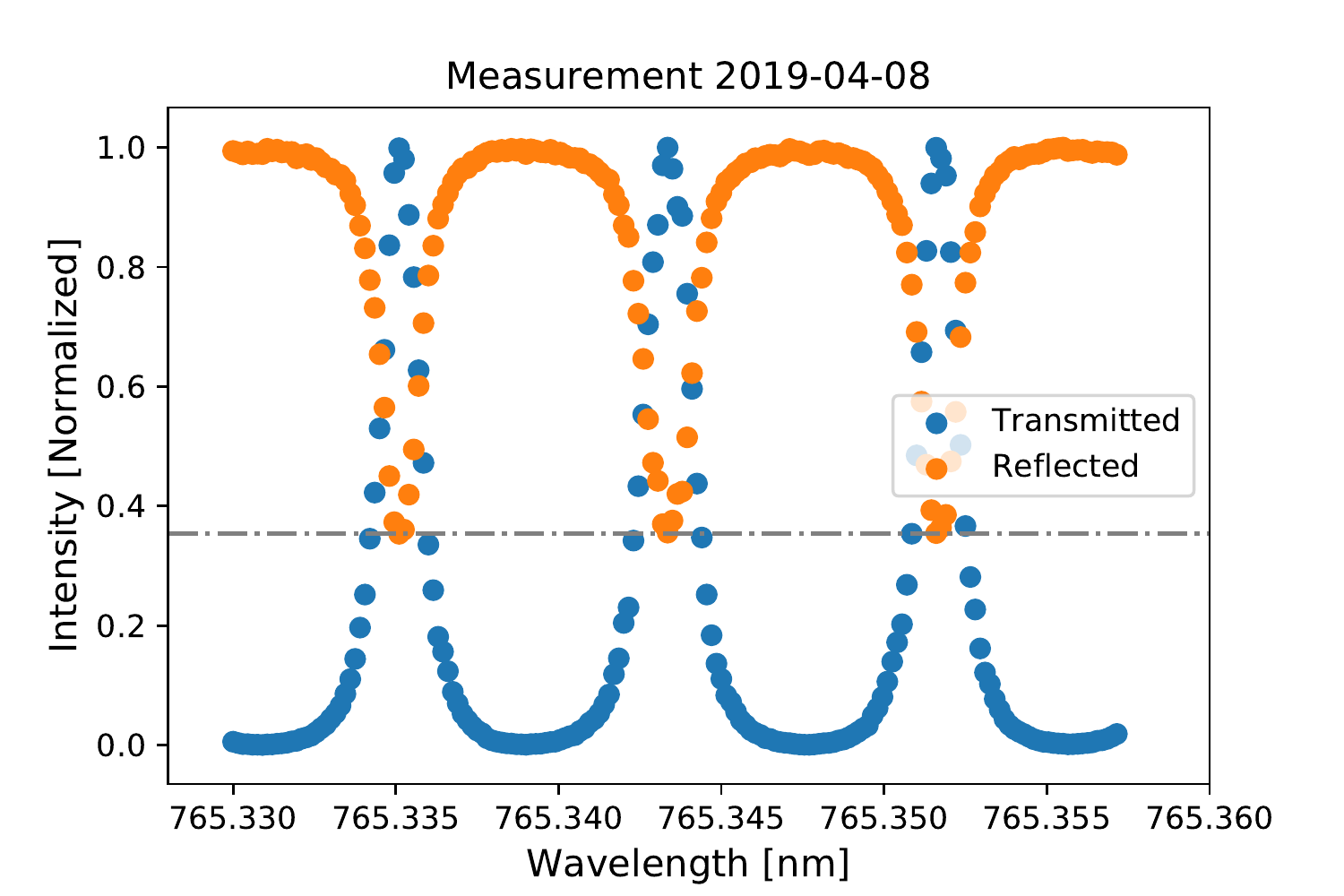}
\includegraphics[width=.47\textwidth]{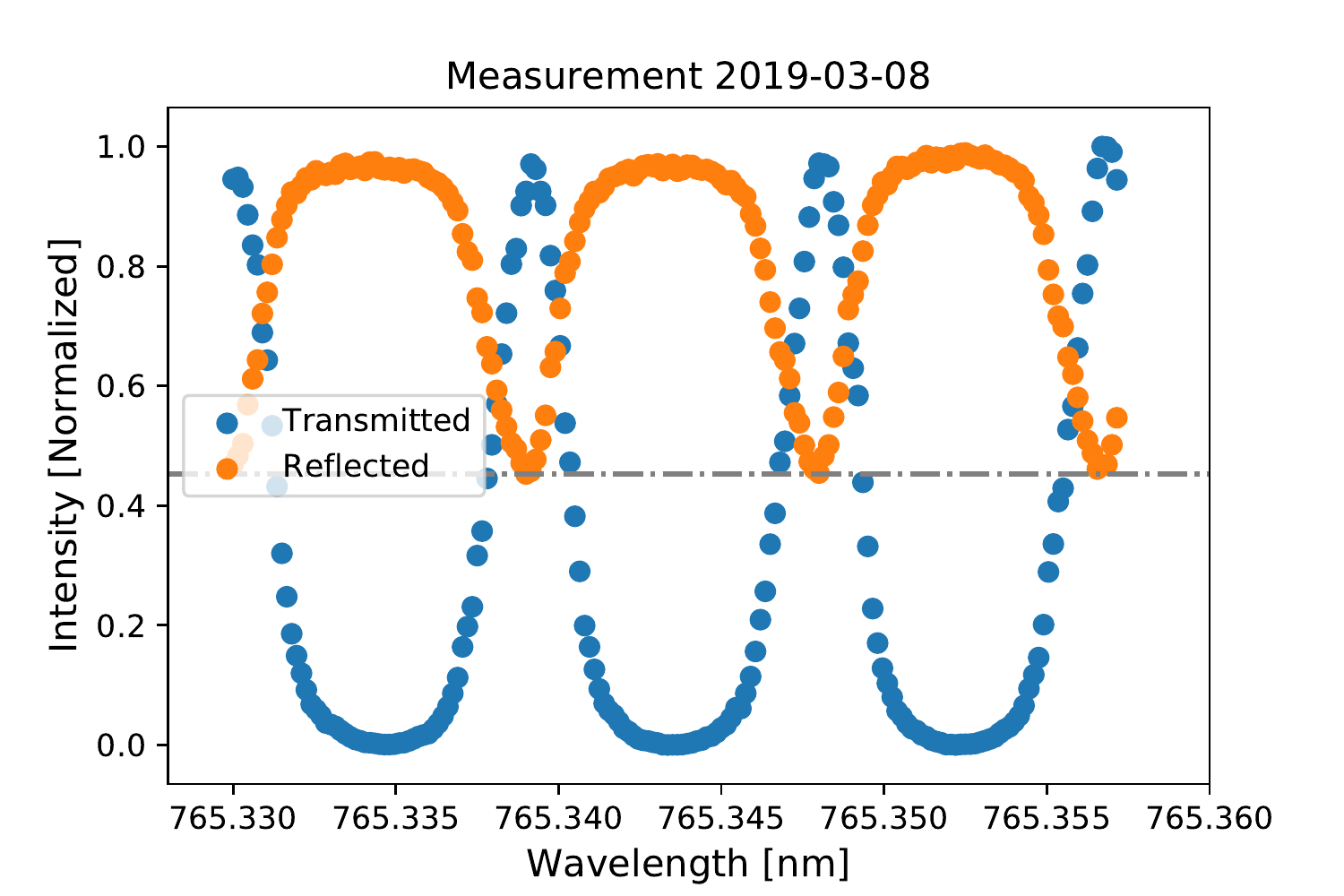}
\caption{Observed data from the setup using a single mode fiber (left) and 50 $\mu$m multi-mode fiber (right). The blue data points are transmitted data from the first arm and the orange ones are the reflected data from the second arm without the second etalon. The dot-dashed line shows the loss level between the first and the second arm.}
\label{fig:observed_TR}
\end{figure}

Figure \ref{fig:observed_TR} is the result obtained from the same setup in Figure \ref{fig:prototype} without the second etalon in the second arm. The two data is fitted using a uniform distribution which accounts for the deviation for each arm. The reflected light from the first arm becomes the incoming light for the second arm according to the Eq. \ref{eq:I_T2}.
The deviation effect is shown as loss in the reflected signal from the second arm. From the plot in Figure \ref{fig:observed_TR}, one can notice at the grey dashed line that the signal from the second arm does not reach the valley of the normalized signal, which means there is loss in the system. We can compare this observation to the predicted profile of the single mode fiber shown in Figure \ref{fig:TRsmf}. Furthermore, we observed that as we increased the fiber size, the loss also increased.

\begin{figure}[h!]
\center
\includegraphics[width=.5\textwidth]{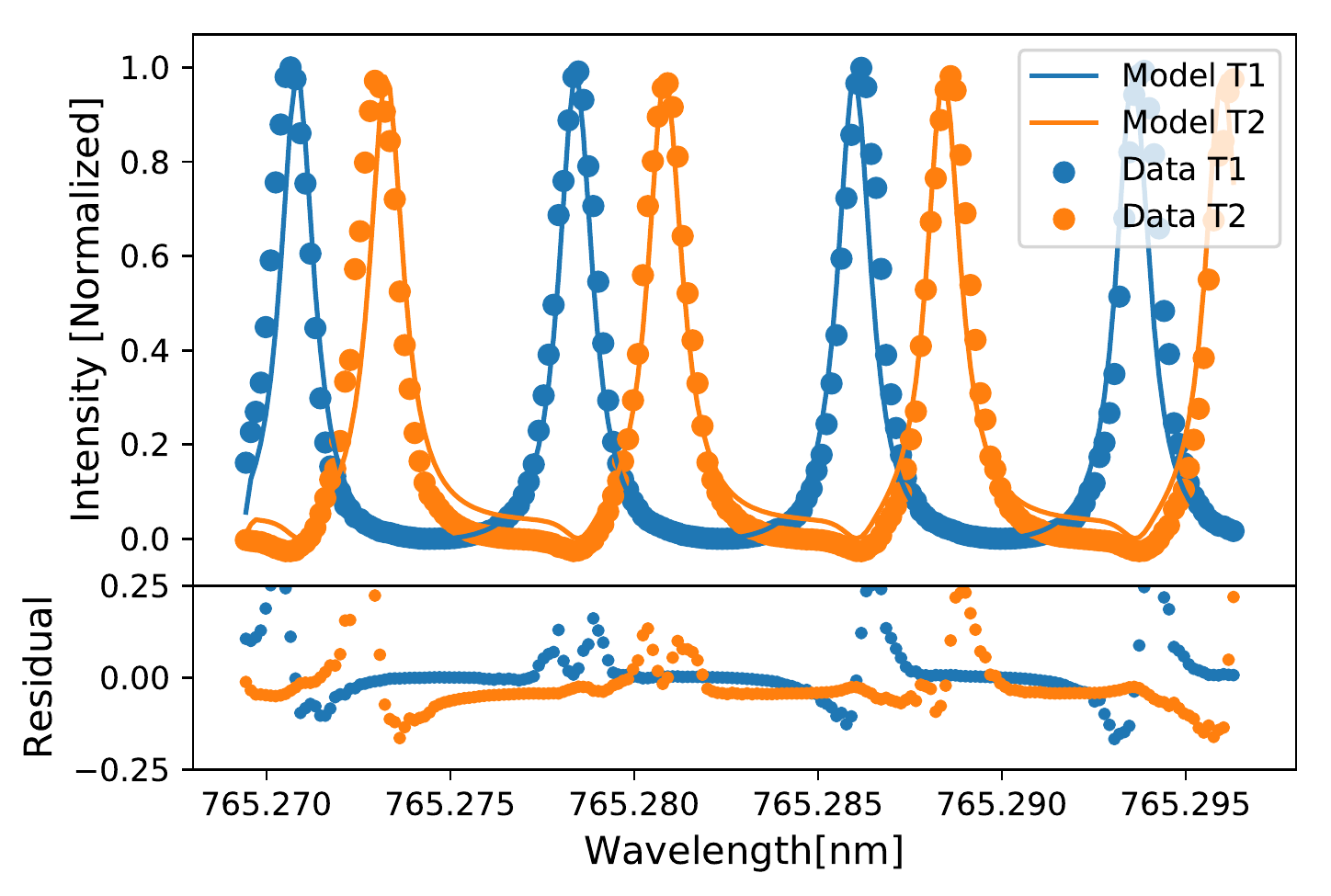}
\caption{The signal from the first arm (blue) and second arm (orange) of the FPI prototype fed by a single mode fiber. The dots are data points and overplotted is a convolved model. Each obtains FWHM 0.0012321 nm, corresponds to the spectral resolution of R = 632,361}
\label{fig:observed1res_etalons}
\end{figure}

Figure \ref{fig:observed1res_etalons} shows the normalized intensity from the single mode fiber where we set the first arm and the second at one resolution element apart. We plot the data points and overlay the best-fit model. The spectral resolution reaches R=704,592.  The current dataset of the single mode fiber is better than the model because we have additional tip-tilt adjustment (tip-tilt stage) in the prototype setup; this parameter is not included in the theoretical model \cite{vaughan}.

\subsection{Dualons}

\begin{figure}[h!]
\includegraphics[width=.45\textwidth]{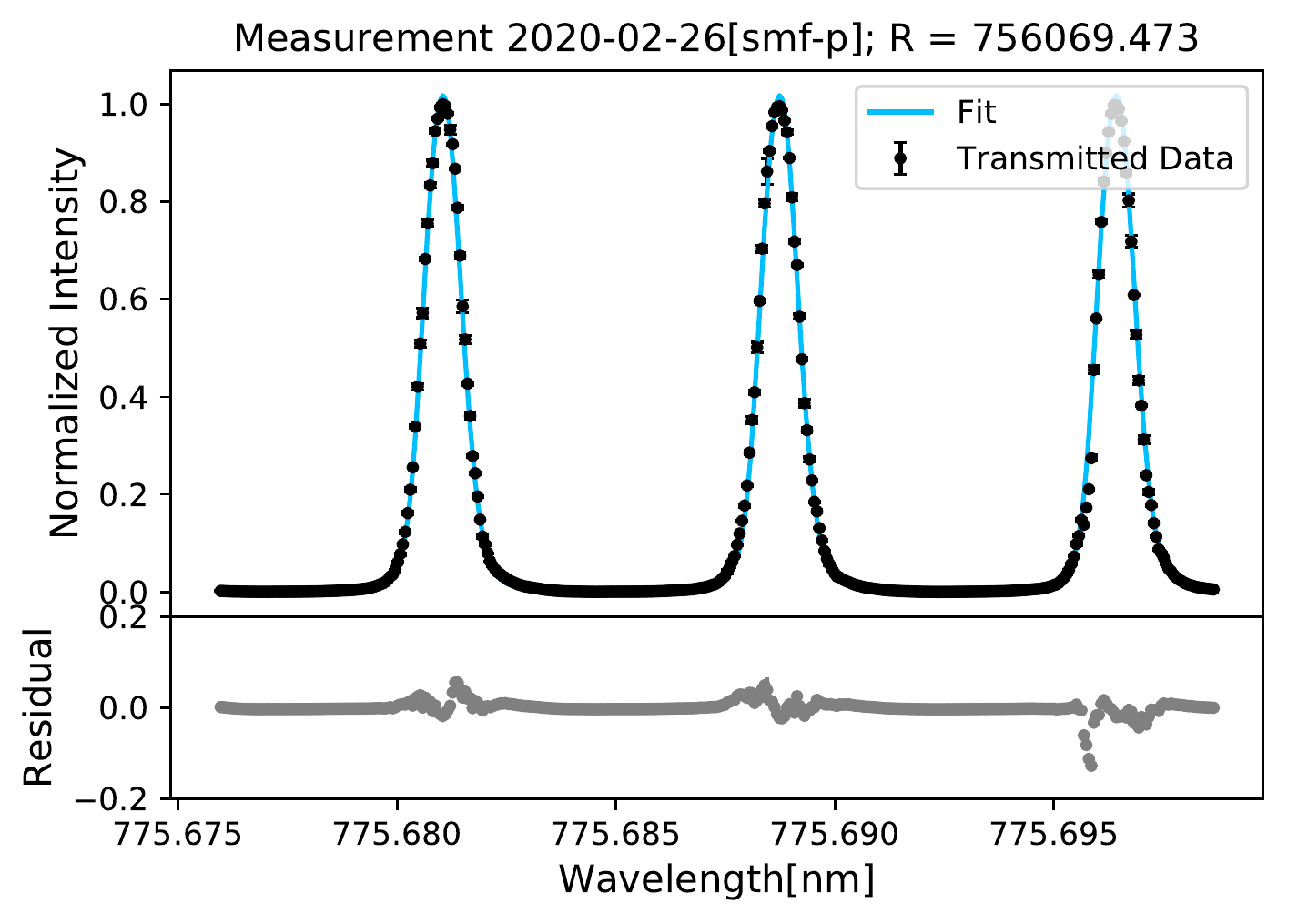}
\hfill
\includegraphics[width=.45\textwidth]{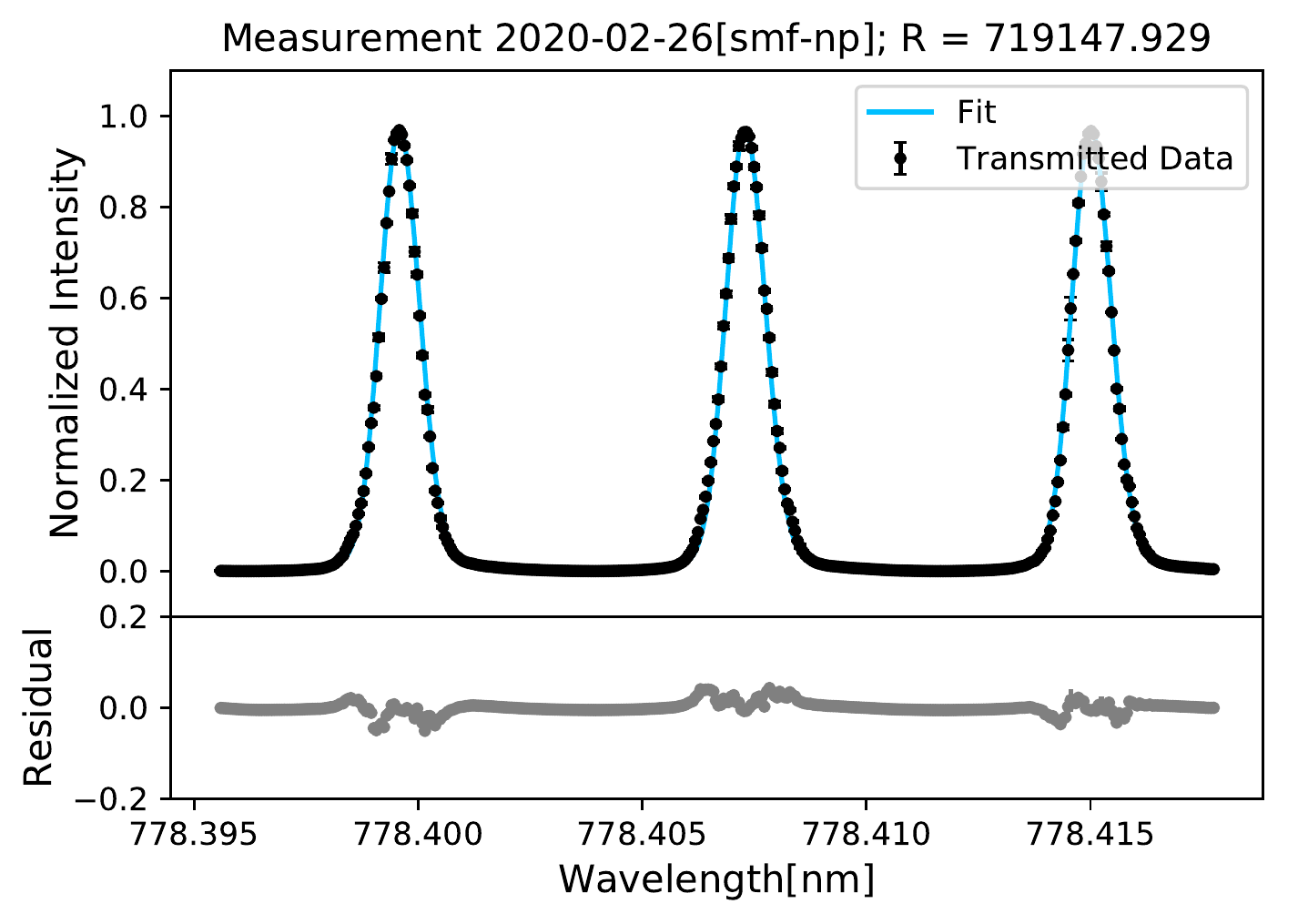}

\includegraphics[width=.45\textwidth]{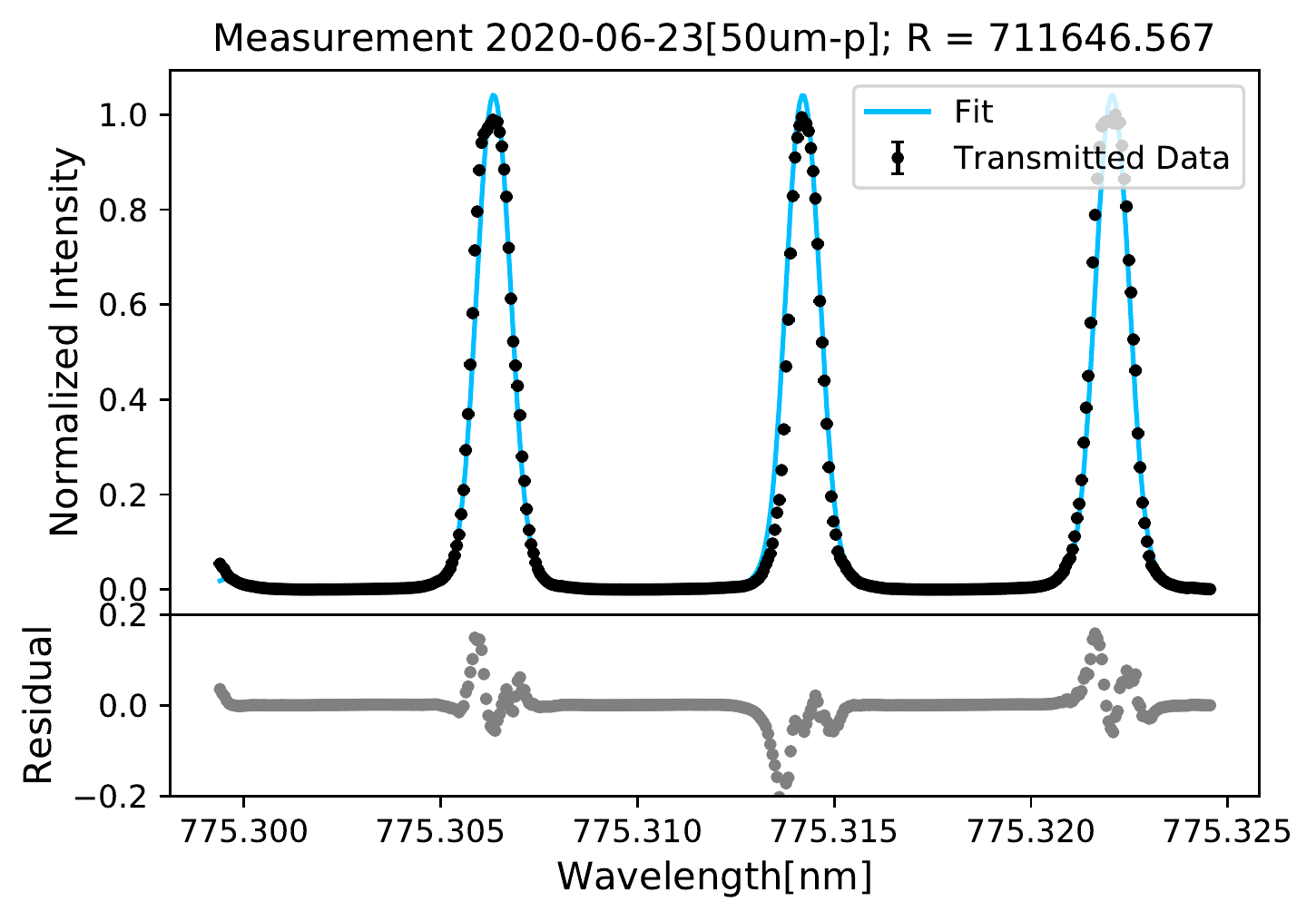}
\hfill
\includegraphics[width=.45\textwidth]{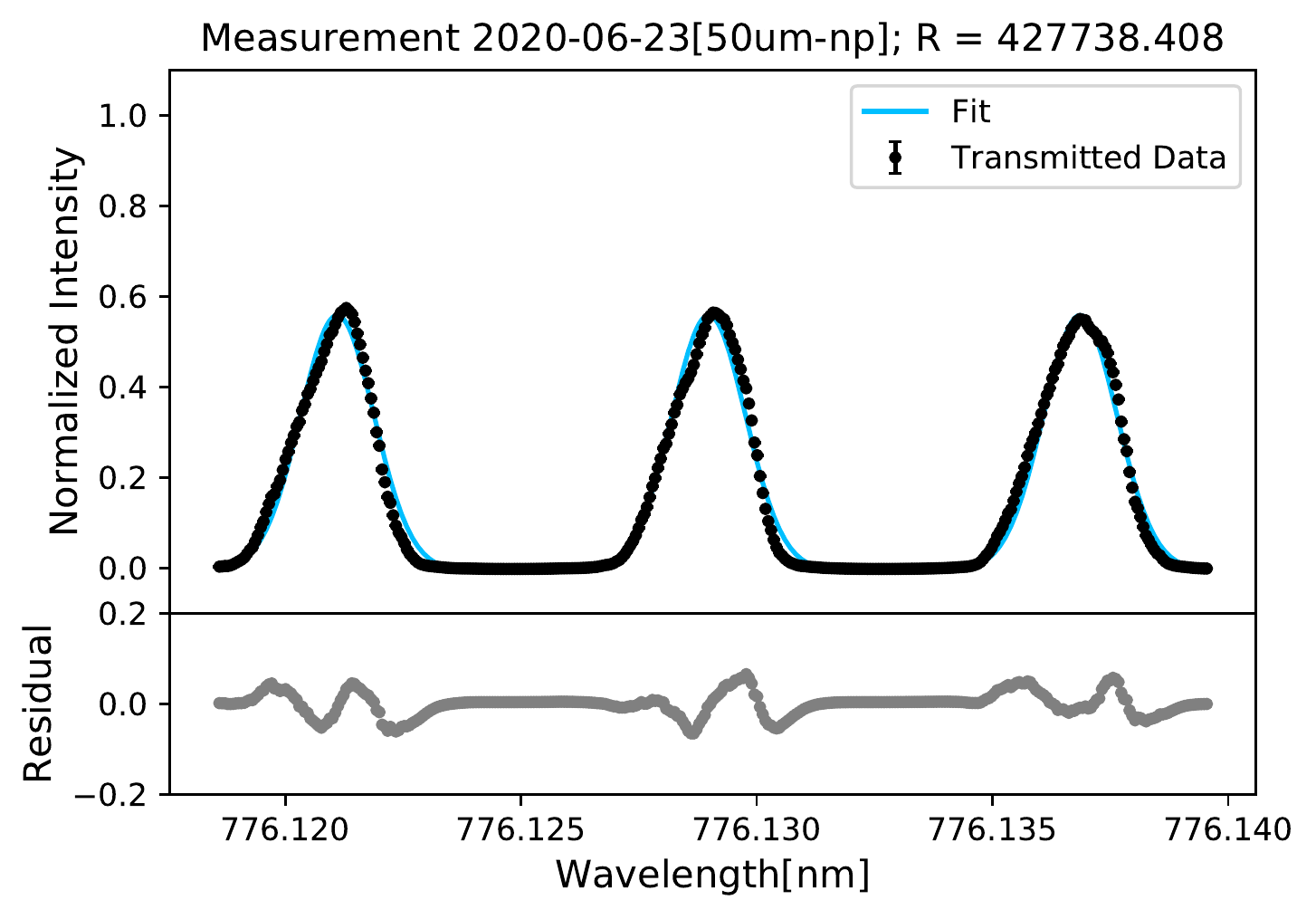}

\includegraphics[width=.45\textwidth]{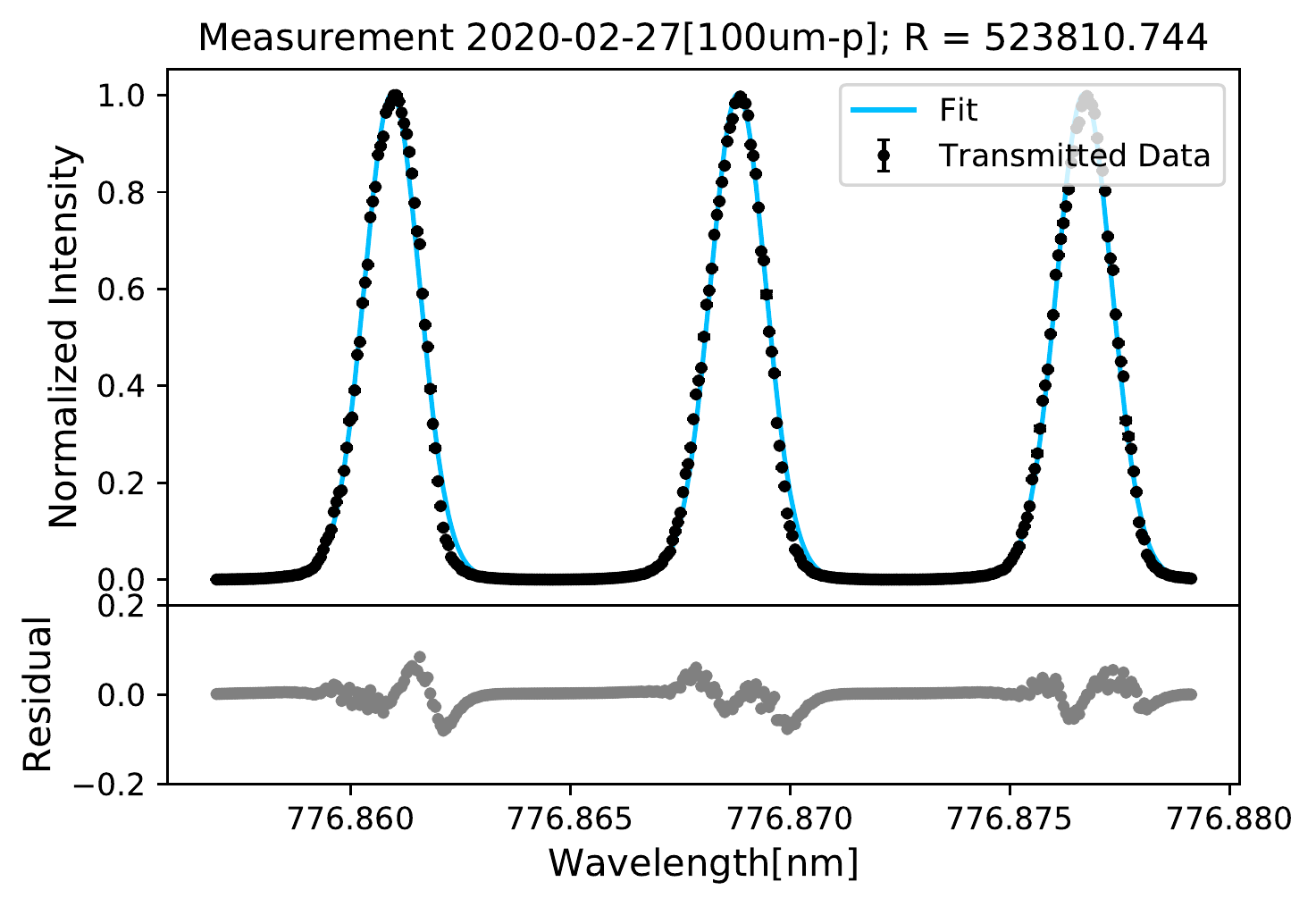}
\hfill
\includegraphics[width=.45\textwidth]{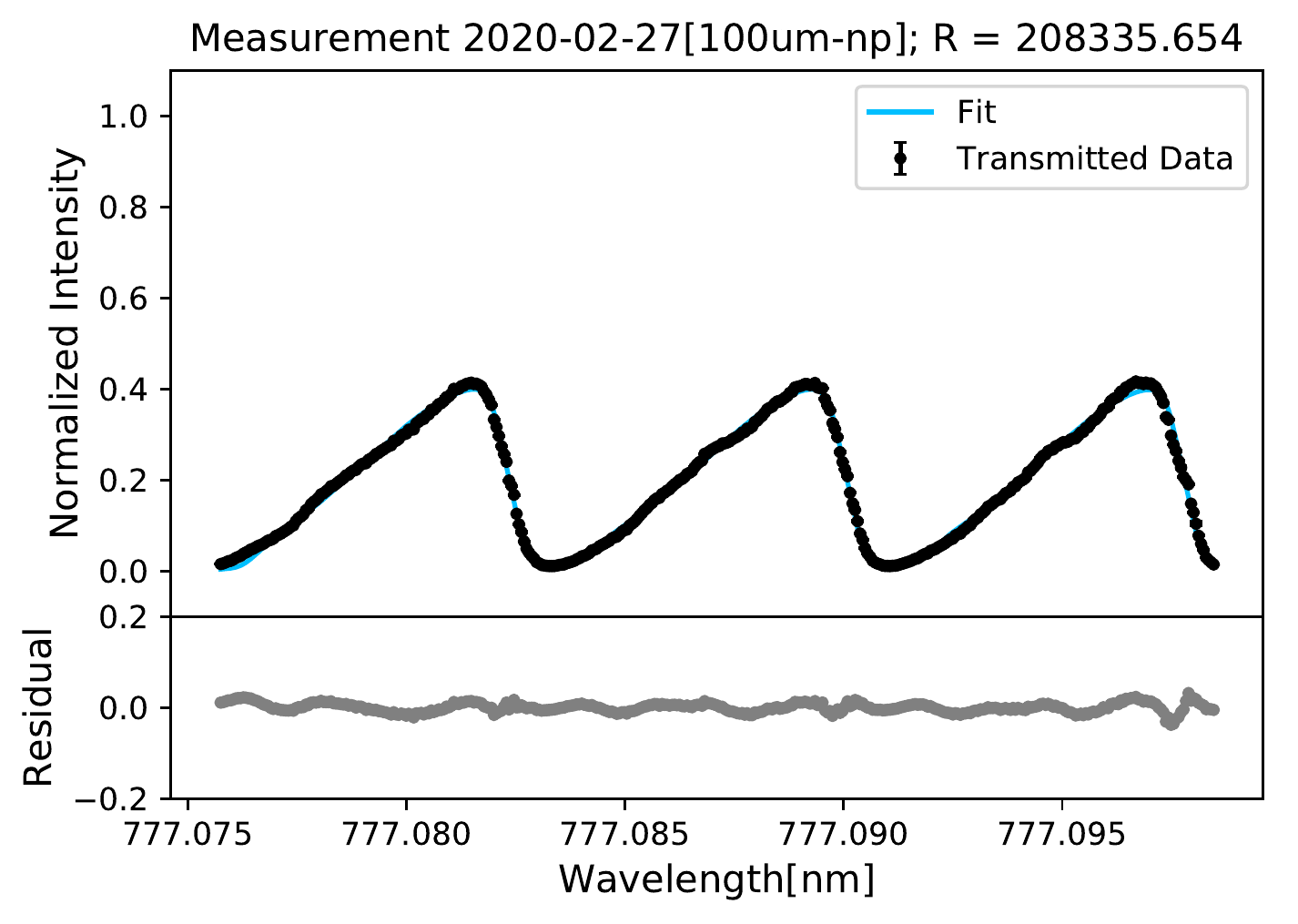}
\caption{The dualon signal from the first arm shows a flat valley. The left panel are data taken in the zero angle position. The right panel are data from the small angle position. From top to bottom row are the variations of the fiber size from single mode fiber, 50$\mu$m, and 100$\mu$m.}
\label{fig:dualon_profile}
\end{figure}

The same measurement setup is applied using dualons instead of etalons. The efficiency of each measurement from the dualon setup is presented in  Table.\ref{tab:Throughput}. The $P_{max}$ is the maximum power of the system measured without placing a dualon in place. This varies from the different size of the fiber similar to the prediction in Fig.\ref{fig:predict}. We compare two major configurations: perpendicular input, where the incident angle is zero (p) and non-perpendicular position (np), where we introduce an angle to the FPI so that the light reflects into the second array. The results are in Fig.\ref{fig:dualon_profile}. 

\begin{table}[h!]
\renewcommand{\thetable}{\arabic{table}}
\centering
\caption{Efficiency Measurement from the dualon setup}  \label{tab:Throughput}
\begin{tabular}{l|r|rr|rr}
\hline
Fiber & $P_{max}$[mW] & p[mW] & loss[\%] & np[mW] & loss[\%]\\
\hline
%\decimals
smf & 10.180 & 10.200 & 0 & 10.050 & 1.277\\
50$\mu$m & 8.700 & 8.750 & 0 & 5.280 & 39.31\\
100$\mu$m & 8.114 & 6.700 & 21.104 & 2.840 & 64.999\\
\hline
\end{tabular}
\end{table}

\begin{figure}[h!]
\center
\includegraphics[width=.5\textwidth]{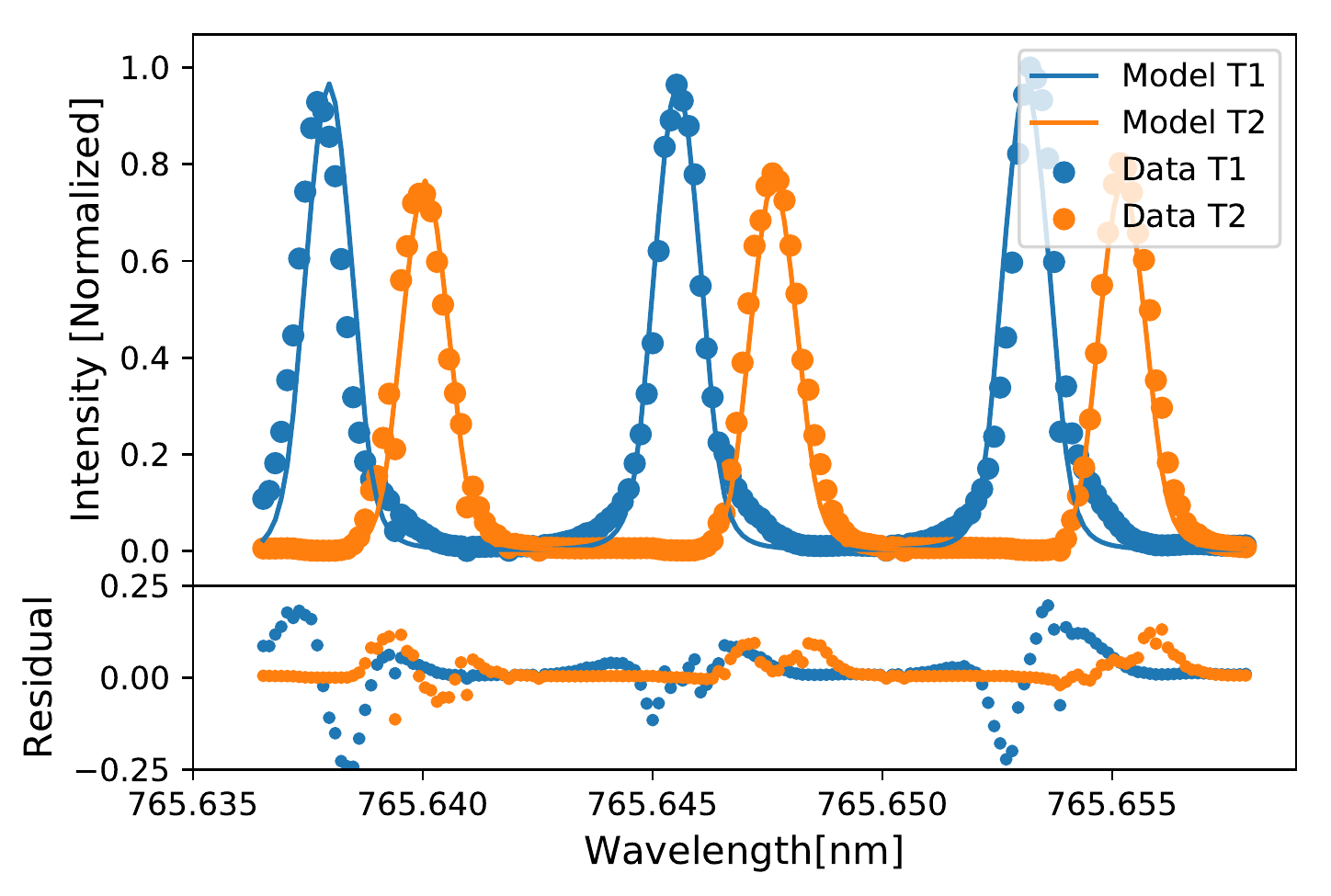}
\caption{The chained spectra observed from the setup using a single mode fiber. The two dualons are tuned to the one resolution element setup. The data points overlay by the fitted uniform function. Blue is the data from the first arm and orange is the second arm. This yields the spectral resolution of R= 663,209 at 765 nm.}
\label{fig:observed1res_dualons}
\end{figure}

\begin{figure}
\center
\includegraphics[width=.75\textwidth]{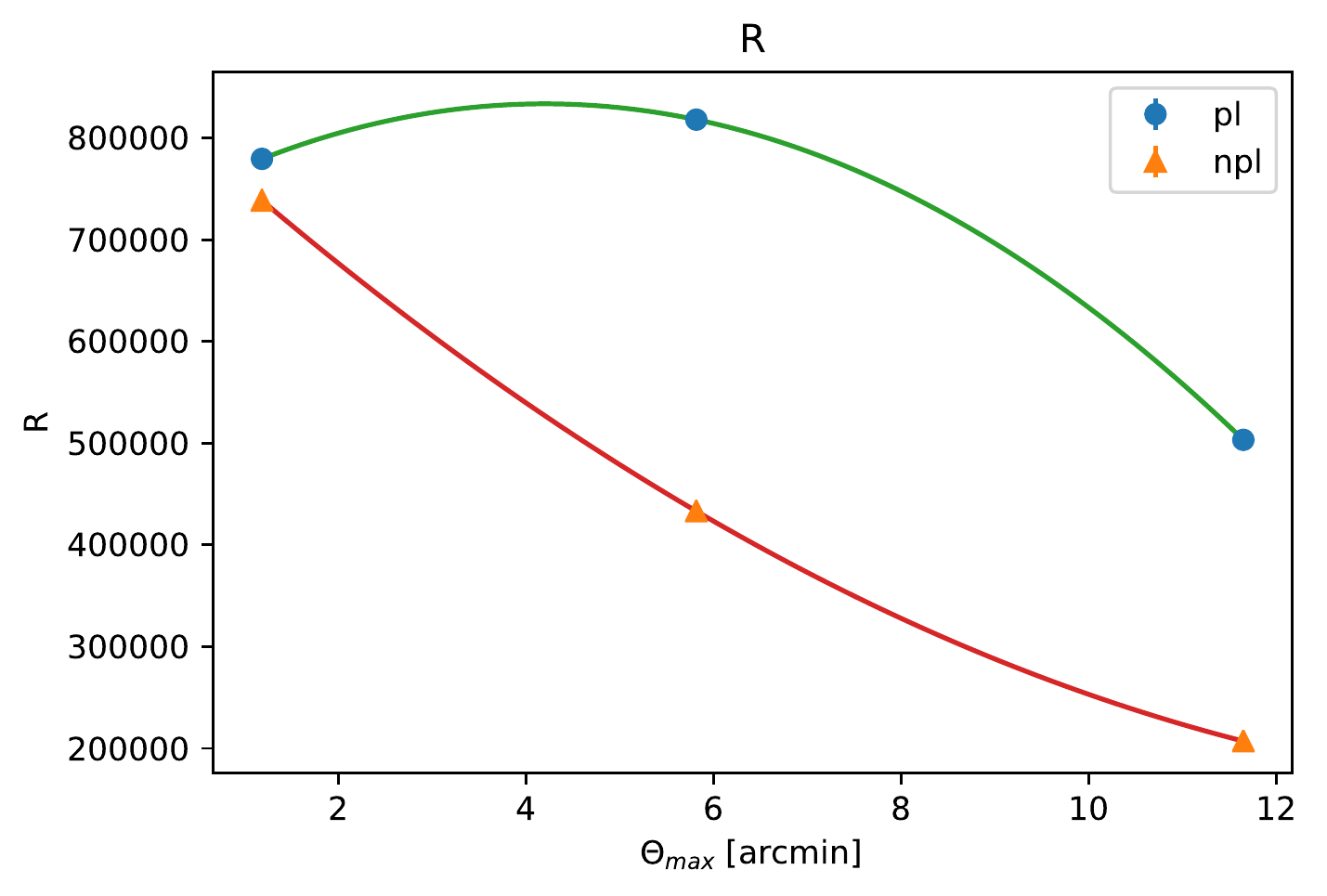}
\caption{The relationship of the measured spectral resolution and the beam deviation. The green line shows a fitted quadratic polynomial result from the perpendicular position and the red line shows the trend from the non-perpendicular configuration.}
\label{fig:Resolution}
\end{figure}

The dualon profiles in Fig.\ref{fig:dualon_profile} are fitted with the multi-mirror equation in Eq.\eqref{eq:I_T_multimirror} and Eq.\eqref{eq:finesse_multimirror} convolved with a skew-normal distribution. The dualon profiles provide a better contrast of the spectral profile. The non-perpendicular position results agree with the prediction (Fig.\ref{fig:predict}) that the larger the fiber size, the lower the efficiency and spectral resolution. This can also be explained by a relationship between the spectral resolution and beam deviation angle shown in Fig.\ref{fig:Resolution}. 

The chained dualon spectra are demonstrated in Fig.\ref{fig:observed1res_dualons}. The transmission profiles are shown in different colors. They are fitted with Eq.\ref{eq:I_T_multimirror} and a Gaussian convolution where the finesse is $F = \frac{4R}{(1 - R)^6}$. The initial incident angle is assigned according the observed size of the beam deviation, which is approximately 5 arcminutes for the single mode fiber. The fitted model reveals the incident angle value to be 5.088 arcminutes for the first arm and 5.406 arcminutes for the second arm. It reaches the spectral resolution of R= 663,209 at 765 nm.

\section{Discussion and Conclusion} \label{sec:discussion}  \label{sec:conclusion}

Previously, we presented (Ref. \citenum{Ben_Ami_2018}) a theoretical model of the etalon array that could be used to finely sample wavelength space and fully characterize the ${\rm O_2}$ A-band (760-780 nm). In this work, we built a two-arm prototype to evaluate the feasibility of their theoretical predictions, comparing their spectral resolution and throughput expectations with empirical measurements through our set-up in the lab. We account for the system imperfection due to the alignment and beam deviation using Eq. \ref{eq:beam deviation}. The expected result from this setup is shown in Figure \ref{fig:predict}. We measured the profile shape, spectral resolution and efficiency. Then we apply the real parameters of the employed etalons and dualons in the model and convolve it with the observed signal level from the setup. The decreasing efficiency due to the plate imperfection shows in our prototype result in Figure \ref{fig:observed_onearm}. In our two-arm implementation, the second arm receives a direct deviation effect from the first arm reflection shown as loss in Figure \ref{fig:TRsmf}, and the alignment imperfection, including plate and collimation imperfection, of its own setup. In other words, the plate imperfection of each arm needs to be considered independently. Further, we found that the different input fiber size and its deviation contribute to the effective Finesse and shape in the second arm. The larger the fiber, the larger the deviation and loss. This applies to both etalon and dualon. In our application (high resolution spectral chain), the setup should be one resolution element apart between each arm transmission peak as an example in Figure \ref{fig:ideal1res}. The observation in the right panel agrees with the prediction in the left panel although the effective Finesse assigned are different. This is because we have additional tip-tilt correction in setup and it helps improving the contrast due to the Finesse especially with the single mode fiber with less deviation. While employing the dualons, the overall contrast has been improved following $F = \frac{4R}{(1 - R)^6}$. We reached spectral resolution of $R=700,000$ with the single mode fiber and $R=400,000$ with multi-mode 50$\mu$m fiber, including the beam deviation, while keeping good contrast of the spectral profile. However, a much larger deviation from the 100$\mu$m profile deconstructs the ideal FPI profile shape completely.

This setup is an implementation to achieve the high resolution spectral train using multi-etalon in a non-traditional position. The main issue we encounter comes from the deviation of the beam. This could be solved by a better ratio between the fiber size, collimator focal length and etalon aperture size. Currently for the highest resolution (R=500,000), we have a setup with a single mode fiber, $f_{col}$ = 30 mm and etalon aperture is 13 mm. In a previous study (Ref.\citenum{2017Cersullo}) for the same resolution using a 200 $\mu$m fiber,  $f_{col}$ = 30 mm and etalon aperture is 40 mm. This implies that using a larger fiber input is possible especially if we want to combine this FPI array with a spectrograph like G-CLEF, which employs a 100 $\mu$m fiber. The aperture size of the optical component in the system needs to be larger than this prototype. The study of Ben-Ami et al. \cite{Ben_Ami_2018} addressed an example application with a giant telescope (e.g. GMT) with a diameter of 25.4 m of the primary mirror. The facet area of the FPI can be calculated from the relation between telescope etendue and beam size on the collimator (Eq. 11\cite{Ben_Ami_2018}). If we aim for the spectral resolution R=500,000, with seeing conditions of 0.7" and telescope primary mirror diameter of 25.4 m, the minimum FPI plate size requirement is 21.55 mm.

%\section{Conclusion}
To sum up, we presented a prototype implementation of the novel Fabry Perot based instrument for characterizing exoplanet atmospheres. Our two-arm instrument prototype incorporates two custom-built FPIs; every other component in our set-up is commercially available. We presented preliminary results demonstrating the achievable spectral sampling and throughput when using either etalons or dualons as FPI elements. Etalons can yield an extremely high resolving power of $\approx 500,000$ at 765 nm, and a maximum throughput of $80\%$. Dualons yield $15\%$ higher spectral resolution than etalons for any fiber size, as well as approximately $10\%$ higher throughput. Future experiments with our prototype include carrying out day-time observations of the Solar spectrum to determine whether we can resolve the telluric oxygen A band.

\acknowledgments % equivalent to \section*{ACKNOWLEDGMENTS}       
This work was made possible through the support of a grant from the John Templeton Foundation. The opinions expressed here are those of the authors and do not necessarily reflect the views of the John Templeton Foundation. We thank the Brinson Foundation and the Smithsonian Institution for providing funding to support this project. JGM is supported by NSF Graduate Research Fellowship Grant No. DGE1745303, and a Ford Foundation Predoctoral Fellowship administered by the National Academies of Sciences, Engineering, and Medicine.

% References
\bibliography{ref} % bibliography data in report.bib
\bibliographystyle{spiebib} % makes bibtex use spiebib.bst

\end{document}